\documentclass[a4paper,11pt]{article}
\pdfoutput=1 

\usepackage{jheppub} 

\usepackage[T1]{fontenc} 
\usepackage{subfigure}

\title{\boldmath Search for the $t\to ch$ decay at hadron colliders}

\author[a,1]{M. A. Arroyo-Ure\~na,}
\author[a]{R. Gait\'an,}
\author[b]{E. A. Herrera-Chac\'on,}
\author[a]{J. H. Montes de Oca Y.}
\author[b]{T. A. Valencia-P\'erez,}

\affiliation[a]{ Departamento  de  F\'isica,  FES-Cuautitl\'an,
Universidad  Nacional  Aut\'onoma  de  M\'exico,
C.P.  54770,  Estado  de  M\'exico,  M\'exico.}
\affiliation[b]{Facultad de Ciencias F\'isico-Matem\'aticas\\
Benem\'erita Universidad Aut\'onoma de Puebla, C.P. 72570, Puebla, Pue., M\'exico.}

\emailAdd{marcofis@yahoo.com.mx}
\emailAdd{rgaitan@unam.mx}
\emailAdd{edwinali@outlook.com}
\emailAdd{josehalim@comunidad.unam.mx}
\emailAdd{antonio.valenciap@alumno.buap.mx}

\abstract{We study the observability for the flavor-changing decay of a top quark $t\to ch$ at the Large Hadron Collider (LHC) and future hadron colliders, namely, High-Luminosity LHC (HL-LHC), High-Energy LHC (HE-LHC) and Future Circular hadron-hadron Collider (FCC-hh). Two scenarios in which the Higgs boson could decay: into a quark bottom pair $(\textit{bb-channel})$ and two photons $(\textit{$\gamma\gamma$-channel})$ are analyzed. A Monte Carlo analysis of the signal and the Standard Model (SM) background is computed. Center-of-mass energies of $\sqrt{s}=14,\;27\; $\text{and}$\; 100$ TeV and integrated luminosities from $0.3$ to $30$ ab$^{-1}$ are explored. The theoretical framework adopted in this work is the Type-III Two-Higgs Doublet Model (THDM-III) for which, constraints on the parameter space from the Higgs boson coupling modifiers $\kappa_i$ are presented and used in order to evaluate the branching ratio of the $t\to ch$ decay and the ($pp\to t\bar{t},\,t\to ch$) production cross section. We find that with the integrated luminosity achieved at the LHC, the  $t\to ch$ decay is out of the reach of detection. More promising results emerge for the HL-LHC, HE-LHC and FCC-hh in which potential discoveries could be claimed.}

\begin{document} 
\maketitle
\flushbottom

\section{Introduction}
The SM is the most successful model to explain almost all the experimental data nowadays. However, despite its great success it is well known that it does not offer adequate answers to some questions such as it does not propose a candidate for dark matter, does not incorporate gravitational interaction, does not give an adequate solution to the hierarchy problem, etc. In particular, in the SM Flavor Changing Neutral Currents (FCNC) mediated by the Higgs boson are not induced at tree-level. The branching ratio for the $t\to ch$ decay in the context of the SM at one-loop level is of the order of $10^{-15}$ \cite{Eilam:1990zc}, \cite{Mele:1998ag}, \cite{AguilarSaavedra:2004wm} which is far from being detected with the current sensitivity of the LHC. However, several models predict the existence of FCNC at the tree level \cite{Arroyo:2013tna}, \cite{Celis:2015ara}, \cite{Botella:2015hoa}, \cite{1901.01304}, \cite{Badziak:2017wxn}, \cite{DiazCruz:2001gf}  and predict branching ratios of up to $10^{-3}$, which opens the possibility that experiments carried out at the LHC or future hadron colliders can be done with high expectation for a detection, namely:

\begin{itemize}
\item High-Luminosity Large Hadron Collider \cite{Apollinari:2017cqg}. The HL-LHC is a new stage of the LHC starting about 2026 to a center-of-mass energy of 14 TeV. The upgrade aims at increasing the integrated luminosity by a factor of ten ($3$ ab$^{-1}$, $\sim$ year 2035) with respect to the final stage of the LHC ($300$ fb$^{-1}$).

\item High-Energy Large Hadron Collider \cite{Benedikt:2018ofy}. The HE-LHC is a possible future project at CERN. The HE-LHC will be a 27 TeV $pp$ collider being developed for the 100 TeV Future Circular Collider. This project is designed to reach up to 12 ab$^{-1}$ which opens a large window for new physics research.

\item Future Circular hadron-hadron Collider \cite{Arkani-Hamed:2015vfh}. The FCC-hh is a future 100 TeV $pp$ hadron collider which will be able to discover rare processes, new interactions up to masses of around 30 TeV and search for a possible substructure of the quarks. Because the great energy and collision rate, billions of Higgs bosons and trillions of top quarks will be produced, this is an unbeatable opportunity to search for the $t\to ch$ decay. The FCC-hh will reach up to an integrated luminosity of 30 ab$^{-1}$ in its final stage.
\end{itemize}
On the other hand, the ATLAS and CMS collaborations \cite{Aaboud:2018pob}, \cite{Sirunyan:2017uae} searched for the $t\to qh$ decay, with $q=u,\,c$, in the $h\to \gamma\gamma$ and $h\to bb$ channels at 7, 8 and 13 TeV, nevertheless they did not found an excess above the background of the SM. The current upper limits for the $t\to ch$ decay by ATLAS collaboration at $\sqrt{s}=13$ TeV corresponding to an integrated luminosity of $36.1$ fb$^{-1}$ are given by:
\begin{eqnarray}
\mathcal{B}(t\to ch)&<&0.16\%, \\ \nonumber
\mathcal{B}(t\to uh)&<&0.19\%,
\end{eqnarray}
while the CMS collaboration at $\sqrt{s}=13$ TeV corresponding to an integrated luminosity of $35.9$ fb$^{-1}$ imposes less restrictive limits given by:
\begin{eqnarray}
\mathcal{B}(t\to ch)&<&0.47\%, \\ \nonumber
\mathcal{B}(t\to uh)&<&0.47\%.
\end{eqnarray}

In theoretical aspect, the prediction of extension models is in the range of $\mathcal{O}(10^{-6})-\mathcal{O}(10^{-3})$ \cite{Li:1993mg}, \cite{Eilam:2001dh}, \cite{Chen:2013qta}, \cite{Yang:2013lpa}, \cite{Azatov:2009na}, \cite{Botella:2015hoa}, \cite{Abbas:2015cua}, \cite{Gaitan:2017tka}.
As far as the simulation is concerned, the authors of ref. \cite{Kao:2011aa} proposed a strategy for the search for $t\to ch$ at the LHC in the framework of the general Two-Higgs Doublet Model which predicts a $\mathcal{B}(t\to ch)\sim\mathcal{O}(10^{-3})$ by using a value for the coupling $htc=\sqrt{m_t m_c}/v\sim 0.006$, the  Cheng-Sher ansatz \cite{Cheng:1987rs}.

In our work, we study the potential discovery about the $t\to ch$ decay within the framework of the Type-III Two-Higgs Doublet Model with four-zero textures (THDM-III). We study $h\to bb$ and $h\to\gamma\gamma$ channels that could appear in collisions as $pp\to t\bar{t}\to  Wb+ch\to\ell\nu b+cXX$ (with $X=b$ for the $\textit{$bb-$channel}$ and $X=\gamma$ for the $\textit{$\gamma\gamma$-channel}$ ) at hadron colliders.

The organization of our work is as follows. In section \ref{SeccionII} we discuss generalities of the THDM-III including the Yukawa interaction Lagrangian written in terms of mass eigenstates as well as the diagonalization of the mass matrix. Section \ref{SeccionIII} is devoted to the constraints on the
relevant model parameter space whose values will be used in our analysis. The section \ref{SeccionIV} is focused on analysis of $pp\to t\bar{t}\to  Wb+ch\to\ell\nu b+cXX$ production cross sections at the LHC, HL-LHC, HE-LHC and FCC-hh. We also present the Monte Carlo analysis of our signal and its SM main background. Finally, conclusions and outlook are presented in section \ref{SeccionV}.

\section{Theoretical framework}\label{SeccionII}
In this section, we give the theoretical framework on which we rely for our research, i.e. THDM-III  with a four-zero texture. We analyze the Yukawa Lagrangian of the THDM-III and obtain the Feynman rules involved in our calculations.
\subsection{Yukawa Lagrangian}
The Yukawa Lagrangian in the THDM-III is given by \cite{Arroyo:2013tna}
\begin{eqnarray}\label{Lagrangiano}
\mathcal{L}_{Y} & = & Y_{1}^{u}\bar{Q}_{L}^{0}\tilde{\Phi}_{1}u_{R}^{0}+Y_{2}^{u}\bar{Q}_{L}^{0}\tilde{\Phi}_{2}u_{R}^{0}+Y_{1}^{d}\bar{Q}_{L}^{0}\Phi_{1}d_{R}^{0}\nonumber \\
 & + & Y_{2}^{d}\bar{Q}_{L}^{0}\Phi_{2}d_{R}^{0}+Y_{1}^{\ell}\bar{L}_{L}^{0}\Phi_{1}\ell_{R}^{0}+Y_{2}^{\ell}\bar{L}_{L}^{0}\Phi_{2}\ell_{R}^{0}+h.c.
\end{eqnarray}
with
\begin{eqnarray}\label{terminosLagrangiano}
Q_{L}^{0} & = & \left(\begin{array}{c}
u_{L}\\ \nonumber
d_{L}
\end{array}\right),\;L^{0}=\left(\begin{array}{c}
\nu_{L}\\
e_{L}
\end{array}\right),\\
\Phi_{1} & = & \left(\begin{array}{c}
\phi_{1}^{+}\\
\phi_{1}^{0}
\end{array}\right),\;\Phi_{2}=\left(\begin{array}{c}
\phi_{2}^{+}\\
\phi_{2}^{0}
\end{array}\right),\\
\tilde{\Phi}_{j} & = & i\sigma_{2}\Phi_{j}^{*}.\nonumber
\end{eqnarray}
Here $\Phi_i$ $(i=1,\,2)$ denotes the Higgs doublets and $Y_i^f$ stands for $3\times3$ Yukawa matrices.
In the Yukawa Lagrangian both Higgs doublets can be coupled to all fermions, so that we would get two Yukawa terms for each doublet.
The physical particles are obtained through a rotation depending on mixing angle $\alpha$, which relates the real part of the $\Phi_i$ doublets with the neutral physical Higgs bosons as follows:
\begin{equation}
\left(\begin{array}{c}
H^{0}\\
h^{0}
\end{array}\right)=\left(\begin{array}{cc}
\cos\alpha & \sin\alpha\\
-\sin\alpha & \cos\alpha
\end{array}\right)\left(\begin{array}{c}
Re\phi_{1}\\
Re\phi_{2}
\end{array}\right),
\end{equation}
whereas the mixing angle $\beta$ transforms the imaginary part of the $\Phi_i$ doublets to the charged and neutral Higgs bosons in the following way:

\begin{equation}
\left(\begin{array}{c}
G^{0}\\
A^{0}
\end{array}\right)=\left(\begin{array}{cc}
\cos\beta & \sin\beta\\
-\sin\beta & \cos\beta
\end{array}\right)\left(\begin{array}{c}
Im\phi_{1}\\
Im\phi_{2}
\end{array}\right),
\end{equation}

\begin{equation}
\left(\begin{array}{c}
G^{\pm}\\
H^{\pm}
\end{array}\right)=\left(\begin{array}{cc}
\cos\beta & \sin\beta\\
-\sin\beta & \cos\beta
\end{array}\right)\left(\begin{array}{c}
\phi_{1}^{\pm}\\
\phi_{2}^{\pm}
\end{array}\right),
\end{equation}
with the angle $\beta$ given by:
\begin{equation}\label{tanbeta}
\tan\beta=\frac{v_2}{v_1}.
\end{equation}
After of the spontaneous symmetry breaking, mass matrices are defined by:
\begin{equation}
M_f=\frac{1}{\sqrt{2}}\left(v_1Y_1^f+v_2Y_2^f\right), f=u, d, \ell.\label{matricesMASA}
\end{equation}
The physical fermion masses are obtained by rotating the matrices of the eq. \ref{matricesMASA} by a bi-unitary transformation $V_f=\mathcal{O}_fP_f$. Then, the diagonalized mass matrices can be written as:

\begin{eqnarray}\label{matricesMASAdiagonal}
\bar{M}_f & = & \frac{1}{\sqrt{2}}V_f\left(v_1Y_1^f+v_2Y_2^f\right)V_f^{\dagger}, \\ \nonumber
& = & \frac{1}{\sqrt{2}}\left(v_1\tilde{Y}_1^f+v_2\tilde{Y}_2^f\right), 
\end{eqnarray}
where $\bar{M_{f}}$ are the diagonalized matrices whose elements are the fermion masses, i.e., $\bar{M_{f}}=\text{Diag}(m_{f_{1}},m_{f_{2}},m_{f_{3}})$. 
$V_{f}$ diagonalizes the mass matrices, although not necessarily it diagonalizes each one Yukawa matrices, which are denoted by $\tilde{Y}_i^f$, with $i=1,\,2$. Therefore, neutral flavor violating Higgs-fermion interactions will be induced. The explicit form of both $\mathcal{O}_f$ and $P_f$ matrices can be consulted in the appendix \ref{AppxAmatrices}.
On the other hand, the mass eigenstates for fermions can be obtained in the following way:
\begin{equation}\label{eigenestadosmasa}
u=V_u^{\dagger}u^0,\;d=V_d^{\dagger}d^0,\;\ell=V_\ell^{\dagger}\ell^0.
\end{equation}
Once the eqs. \ref{terminosLagrangiano} - \ref{tanbeta} and \ref{eigenestadosmasa} are introduced in the eq. \ref{Lagrangiano}, the $h\bar{t}c$ coupling acquires a very simple form \cite{hep-ph/0401194}, \cite{0902.4490}:
\begin{eqnarray}\label{LagrangianoInteracciones}
\mathcal{L}_{Y} & = &  \bar{t}\left[\frac{\cos(\alpha-\beta)}{\sqrt{2}\sin\beta}\left(\tilde{Y}_{2}^{u}\right)_{ij}\right]ch.
\end{eqnarray}
The complete Yukawa Lagrangian is shown in the appendix \ref{AppxALagrangian}.
We observe that eq. \ref{LagrangianoInteracciones} includes FCNC at tree-level. In order to suppress them, we assume that the Yukawa matrices of the eq. \ref{matricesMASA} have the form of an hermitian four-zero texture, i.e.,
\begin{equation}\label{Yuk4Zero}
Y^{f}_i=\left(\begin{array}{ccc}
0 & d_i^f & 0\\
d_i^{f^*} & c_i^f & b_i^f\\
0 & b_i^{f^*} & a_i^f
\end{array}\right),
\end{equation}
whose elements have the hierarchy: $|a_i^f|\gg |b_i^f|,\,|c_i^f|,\,|d_i^f|$.
Given the structure of the Yukawa matrices as above, the mass matrix inherits its form, so that:
\begin{equation}\label{Matrixmass}
M^{f}=\left(\begin{array}{ccc}
0 & D^{f} & 0\\
D^{f^*} & C^f & B^f\\
0 & B^{f^*} & A^f
\end{array}\right).
\end{equation}
The elements of a real matrix of the type \ref{Matrixmass} are related to eigenvalues $m_i$, ($i=1,\,2,\,3$), through the following invariants:
\begin{eqnarray}\label{Invariantes}
det\left(M\right) & = & -D^{2}A=m_{1}m_{2}m_{3},\nonumber \\
Tr\left(M\right) & = & C+A=m_{1}+m_{2}+m_{3},\\
\lambda\left(M\right) & = & CA-D^{2}-B=m_{1}m_{2}+m_{1}m_{3}+m_{2}m_{3},\nonumber
\end{eqnarray}
where we omit the index $f$, as of now, so as not to overload the notation.
We assume the hierarchy $m_3>A>m_2>m_1$, with $A=m_3-\gamma m_2$ and $\gamma$ in the interval $[0,1]$.
From these expressions we find a relation between the components of the four-zero matrix mass and the mass eigenstates, namely:
\begin{eqnarray}\label{ElementosMatrizMasa}
A & = & m_{3}(1-r_{2}\gamma),\nonumber \\
B & = & m_{3}\sqrt{\frac{r_{2}\gamma(r_{2}\gamma+r_{1}-1)(r_{2}\gamma+r_{2}-1)}{1-r_{2}\gamma},}\\
C & = & m_{3}(r_{2}\gamma+r_{1}+r_{2}),\nonumber \\
D & = & \sqrt{\frac{m_{1}m_{2}}{1-r_{2}\gamma}},\nonumber
\end{eqnarray}
with $r_i=m_i/m_3$.

By considering the eqs. \ref{matricesMASAdiagonal} and \ref{Yuk4Zero} - \ref{ElementosMatrizMasa}, the terms $\left(\tilde{Y}_{2}^{f}\right)_{ij}$ of the eq. \ref{LagrangianoInteracciones} can be written as:
\begin{equation}
\left(\tilde{Y}_{2}^{f}\right)_{ij}=\frac{\sqrt{m_im_j}}{v}\chi_{ij},
\end{equation}
i.e., the Cheng-Sher ansatz multiplied by a term depending on Yukawa matrix elements and phases coming from eqs. \ref{O} and \ref{fases}. In particular, we have:
\begin{equation}
\left(\tilde{Y}_{2}^{u}\right)_{tc}=\frac{\sqrt{m_tm_c}}{v}\chi_{tc},
\end{equation}
where
\begin{equation}\label{chi}
\chi_{tc}=\sqrt{\frac{m_t}{m_c}}\left(\frac{b_1v}{m_t\tan\beta}-\frac{F_1}{\sin\beta}\right)e^{i\alpha_2}+\left(\frac{(a_1-c_1)v}{m_t\tan\beta}-\sqrt{2}\frac{F_2}{\sin\beta}\right)\sqrt{\gamma_u},
\end{equation}
we define, $F_1=\sqrt{2GR}$, $F_2=Q-2G$, $G=r_c\gamma_u$, $R=1-r_c(1-\gamma_u)$, $Q=1-r_c$ and $r_c=m_c/m_t.$
In this work, instead of constraining the parameters that come from the explicit form of Yukawa matrices, we restrict the $\chi_{tc}$ parameter as a whole.

\section{Model parameter space}\label{SeccionIII}
In order to evaluate the decay width and the $(pp\to t\bar{t},\,t\to hc)$ production cross section, it is necessary to have current bounds on the model parameters involved in our calculation.
These free model parameters are the following:
\begin{itemize}
\item $\cos(\alpha-\beta)= c_{\alpha\beta}$,
\item $\tan\beta=t_{\beta}$,
\item $\chi_{tc}$.
\end{itemize}
\subsection{Constraint on $c_{\alpha\beta}$}
\begin{figure}[!htb]
\centering
 \subfigure[ ]{\includegraphics[scale = 0.3]{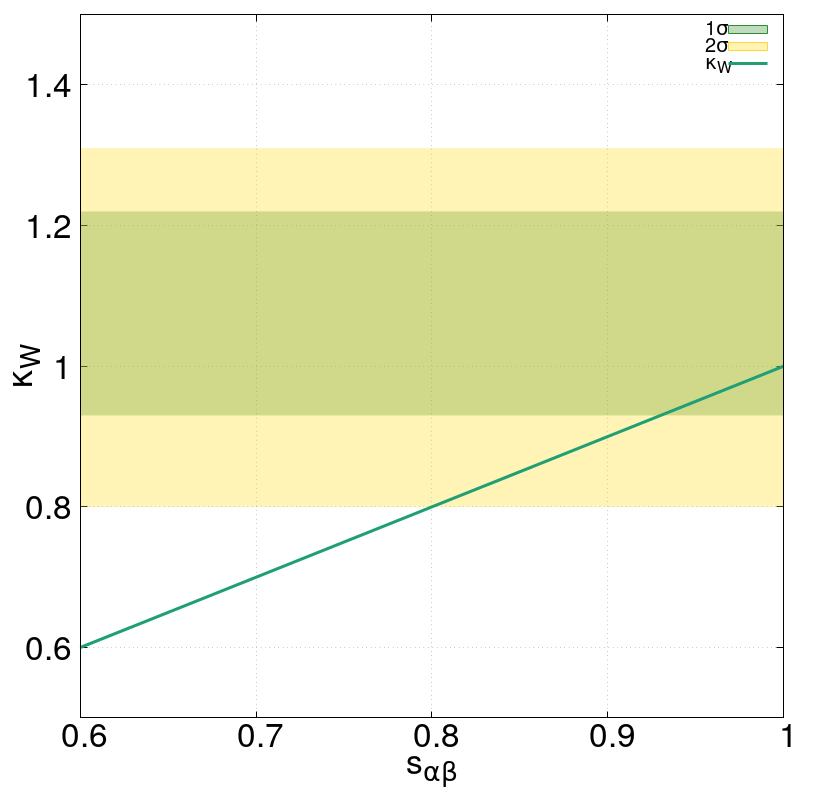}\label{a}}
 \subfigure[ ]{\includegraphics[scale = 0.3]{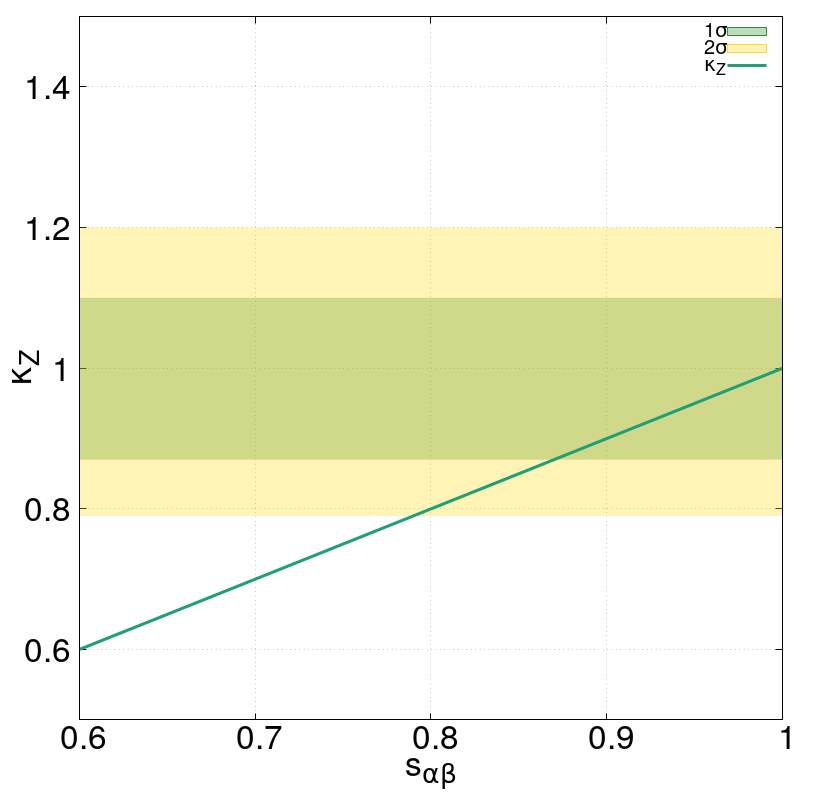}\label{b}}
 \caption{(a) $\kappa_W$ and (b) $\kappa_Z$ coupling modifiers as a function of $s_{\alpha\beta}$. The dark areas represent the allowed regions by the experimental constraints at 1$\sigma$ and 2$\sigma$.    \label{KappaVPlots}}
	\end{figure}

To constrain $c_{\alpha\beta}$, we use the most up-to-date constraints on the Higgs boson data reported by CMS collaboration \cite{Sirunyan:2018koj}:
\begin{itemize}
 \item The Higgs boson coupling modifiers $\kappa_j$ which, for a production cross section or a decay mode $j$, are defined as:
 \begin{equation}
 \kappa^2_j=\sigma_j/\sigma^{\text{SM}}_j\;\text{or}\; \kappa^2_j=\Gamma_j/\Gamma^{\text{SM}}_j.
\end{equation}
\end{itemize}
Effects of new physics arise through $\sigma_j$ and $\Gamma_j$.
Because the $hVV$ coupling coming from THDM-III ($g_{hVV}^{\text{THDM-III}}=\sin(\alpha-\beta)g_{hVV}^{\text{SM}}$, with $V=W,\,Z$) depends on $\sin(\alpha-\beta)=s_{\alpha\beta}$, we use the $\kappa_V$ in order to constrain $c_{\alpha\beta}$. The table \ref{kaV} shows the most up-to-date values for $\kappa_V$ reported by CMS Collaboration \cite{Sirunyan:2018koj}. In the figure \ref{KappaVPlots} is presented the allowed region by $\kappa_V$ in the $s_{\alpha\beta}$-$\kappa_{W(Z)}$ planes. The graphics were obtained through the $\texttt{SpaceMath}$ package \cite{work-in-progress}.
\begin{table}
\caption{The best fit values and $\pm1\sigma$ uncertainties
for $\kappa_{V}$.}\label{kaV}
\begin{centering}
\begin{tabular}{cc}
\hline
Parameter & The best fit value\tabularnewline
\hline
\hline
$\kappa_{W}$ & $1.10_{-0.17}^{+0.12}$\tabularnewline
\hline
$\kappa_{Z}$ & $0.99_{-0.12}^{+0.11}$\tabularnewline
\hline
\end{tabular}
\par\end{centering}
\end{table}
To $2\sigma$, $\kappa_W$ and $\kappa_Z$ impose a low limit for $s_{\alpha\beta}\sim 0.8$, however, by considering $1\sigma$ uncertainties for $\kappa_W$, its lowest limit ($s_{\alpha\beta}\sim 0.93$) is more restrictive than $\kappa_Z$ ($s_{\alpha\beta}\sim 0.86$). We note that in the special case when $s_{\alpha\beta}=1$, then $\kappa_V=1$ and the SM is recovered. Because $h$ is identified with the SM-like Higgs boson, to have a consistent theoretical framework with the SM, we consider $s_{\alpha\beta}=0.99$, which implies that $c_{\alpha\beta}\sim0.14$. These results are in accordance with the analysis reported in the ref. \cite{Carmi:2012in}, in which $(\alpha-\beta)\sim\pi/2$ it is the most favorable scenario. 
\subsection{Constraint on $t_{\beta}$ and $\chi_{tc}$}
In addition to $c_{\alpha}$, also $t_{\beta}$ and $\chi_{tc}$ are free parameters. To constraint them, we consider the direct upper bound on the $\mathcal{B}(t\to tc)<0.16\%$ imposed by ATLAS collaboration \cite{Aaboud:2018pob}, however, with this upper bound a very weak bounds on $t_{\beta}$ and $\chi_{tc}$ are obtained. Nevertheless, the authors of the ref. \cite{Papaefstathiou:2017xuv} have obtained a better upper limit than ATLAS, extrapolating the number of events for the signal and backgrounds from $36.1$ fb$^{-1}$ to $3000$ fb$^{-1}$, assuming that the experimental details and analysis remain unchanged. This upper limit is given by $\mathcal{B}(t\to tc)<0.00769\%$.

\begin{figure}[!htb]
\centering
{\includegraphics[scale = 0.3, angle=270]{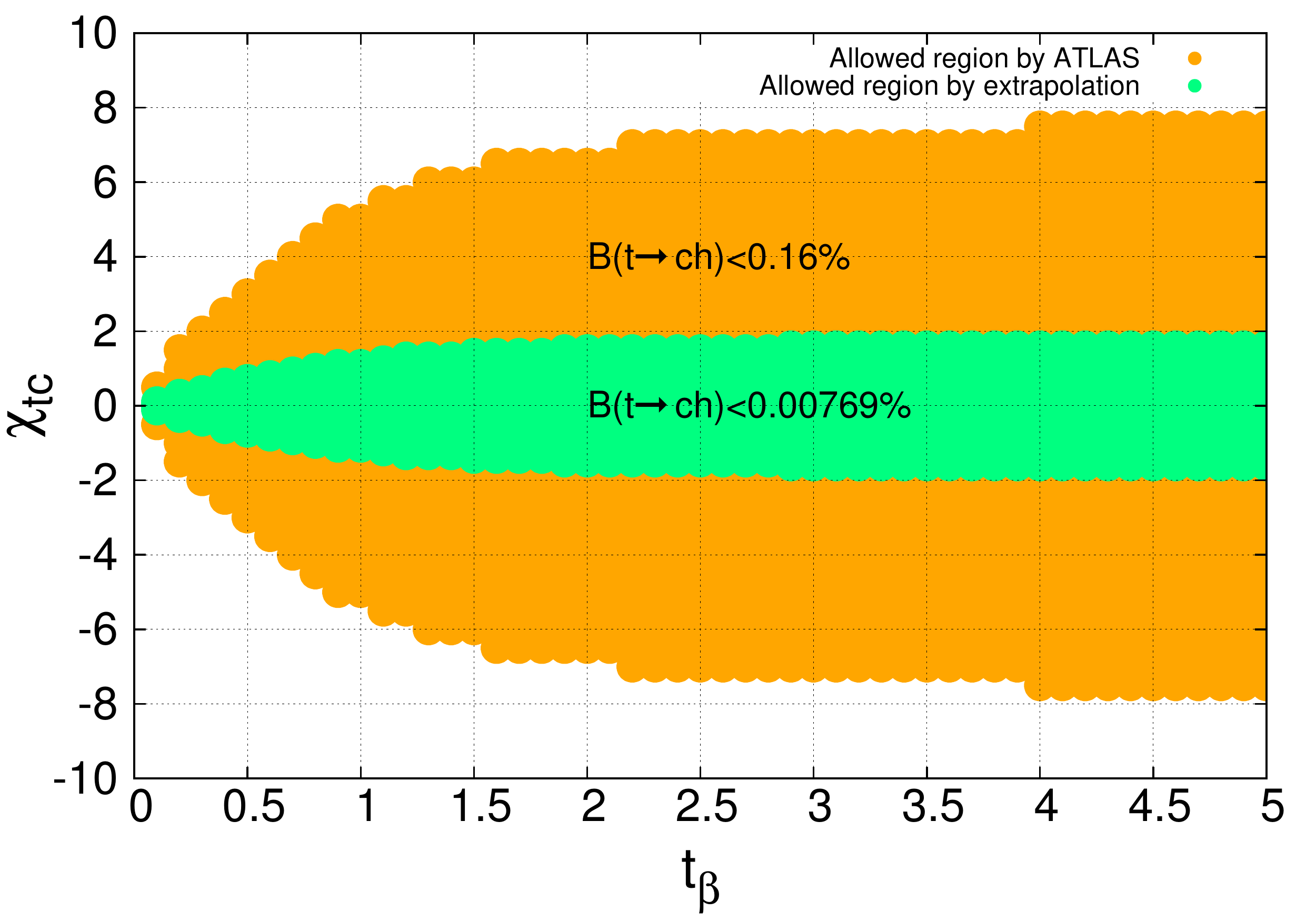}}
 \caption{Allowed region by the reported upper limit by ATLAS (orange area) and by extrapolation (green area). \label{TanBETAvsCHItclabel}}
	\end{figure}

In the figure \ref{TanBETAvsCHItclabel} is presented the allowed region in the $t_{\beta}-\chi_{tc}$ plane by the direct upper bound on the $\mathcal{B}(t\to tc)<0.16\%$ and by extrapolation $\mathcal{B}(t\to tc)<0.00769\%$.
Considering the limit by ATLAS, the allowed values for $\chi_{tc}$ are in the range from $-8$ to $8$ once the $t_{\beta}\sim 4$, whereas for $t_{\beta}\leq 1.5$, $\chi_{tc}$ decreases. On the other hand, if the extrapolation is applied, there will be a more restrictive scenario contemplating values for $ \chi_{tc} $ in the range from $ \sim -2 $ to $\sim 2$ for $1.5\leq t_{\beta}$. In order to get $\tilde{Y}_2^u\sim$ Cheng-Sher ansatz, values for $\chi_{tc}$ between $0.5-1.5 $ are considered, corresponding to values for $t_{\beta}$ in the $(0-1)$ interval. However, the authors of \cite{Babu:2018uik} proposed a ansatz modified for a scalar-fermion interaction. In summary, the table \ref{parametros} presents the values for the free model parameters used in this work.
\begin{table}[!htb]
\caption{Values for the free model parameters used in this work. }\label{parametros}
\begin{centering}
\begin{tabular}{cc}
\hline
Parameter & Values\tabularnewline
\hline
\hline
$c_{\alpha\beta}$ & $0.14$\tabularnewline
\hline
$t_{\beta}$ & $0.1-1$\tabularnewline
\hline
$\chi_{tc}$ & $0.5-1.5$\tabularnewline
\hline
\end{tabular}
\par\end{centering}
\end{table}

\section{Search for $t\to ch$ decay at hadron colliders \label{SeccionIV}}
The main interest in this paper is to study an evidence or a possible discovery of the $t\to c h$ decay. The theoretical framework adopted to study the signal is the THDM-III. The analysis is carried out for the LHC and future hadron colliders:
\begin{enumerate}
\item High-Luminosity LHC \cite{Apollinari:2017cqg}.
\item High-Energy LHC \cite{Benedikt:2018ofy}.
\item Future Circular Collider-hh \cite{Arkani-Hamed:2015vfh}.
\end{enumerate}
In this work two channels are explored, namely, the Higgs boson decaying into two photons (\textit{$\gamma\gamma$-channel}) and two bottom quarks (\textit{$bb$-channel}). Then, the signal and the SM main background processes are as follows:
\begin{itemize}
\item \textbf{SIGNAL:}\\
The signal is $pp\to t\bar{t}\to hc+Wb\to XX c+\ell\nu_{\ell} b$, with $X=\gamma$ for the \textit{$\gamma\gamma$-channel} and $X=b$ for the \textit{$bb$-channel}. Then, final state of the signal is $\gamma\gamma bj\ell\nu_{\ell}$ or $b\bar{b}bj\ell\nu_{\ell}$. The flavor-changing process come from one top quark decaying into a charm quark and a Higgs boson through the production mechanism of top quark pairs.
\end{itemize}
\begin{itemize}
\item \textbf{BACKGROUND:}
\end{itemize}
\begin{enumerate}
\item \textbf{\textit{$\gamma\gamma$-channel:}} Considering the main background processes that include a Higgs boson in association with other particles and non-resonant production of photon pairs:
\begin{itemize}
\item $pp\to t\bar{t}h$,
\item $pp\to hjjW^{\pm}$,
\item $pp\to t\bar{t}\gamma\gamma$,
\item $pp\to\gamma\gamma jjW^{\pm}$.
\end{itemize}
\item \textbf{\textit{$bb$-channel:}} The SM dominant background to the final state $b\bar{b}bj\ell\nu$ are as follows:
\begin{itemize}
\item $pp\to t\bar{t}\to b\ell^+\nu\bar{b}\bar{c}s+X$ or $pp\to t\bar{t}\to bc\bar{s}\bar{b}\ell^-\bar{\nu}+ X$, with a $c$-jet is mis-identified as a $b$-jet,
\item $pp\to t\bar{t}\to b\ell\nu\bar{b}u\bar{d}$,
\item $pp\to b\bar{b}b\bar{b}\ell\nu$,
\item $pp\to b\bar{b}c\bar{c}\ell\nu$.
\end{itemize}
\end{enumerate}
\subsection{Number of signal events}

We now turn to analyze the number of events produced for the signal as a function of $t_{\beta}$ and $\chi_{tc}$ at the LHC and future hadron colliders, i.e., HL-LHC, HE-LHC and FCC-hh. We consider events if and only if they satisfy the constraint $\mathcal{B}(t\to ch)<10^{-5}$, i.e., two orders of magnitude less than the upper limit reported by the ATLAS  \cite{Aaboud:2018pob} and CMS \cite{Sirunyan:2017uae} collaborations and slightly more restrictive than the one reported in ref. \cite{Papaefstathiou:2017xuv}.
\subsubsection{$\gamma\gamma$-channel}
\begin{figure}[!h]
\centering
\subfigure[]{\includegraphics[scale = 0.25,angle=270]{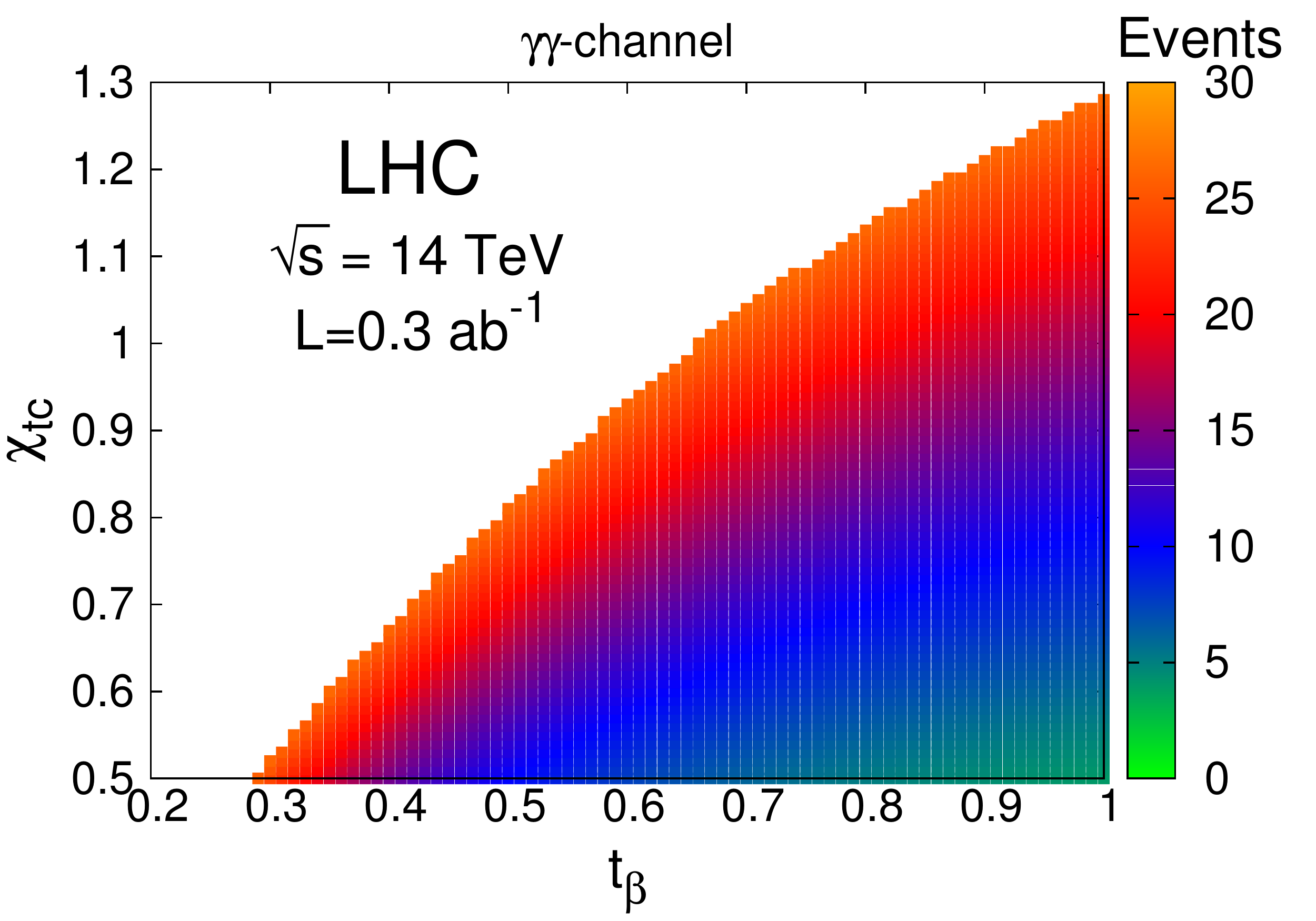}}
\subfigure[]{\includegraphics[scale = 0.25,angle=270]{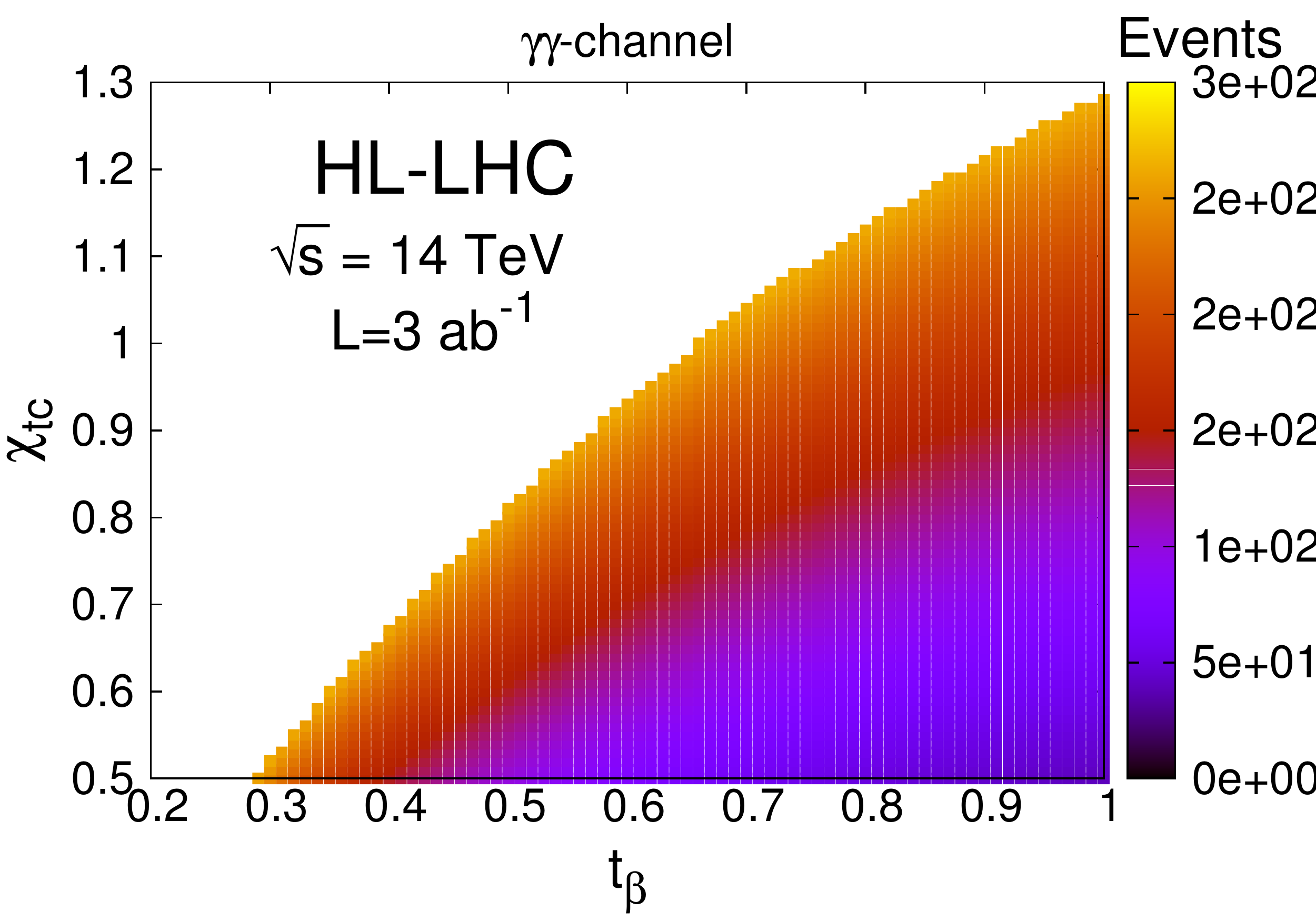}}
\subfigure[]{\includegraphics[scale = 0.25,angle=270]{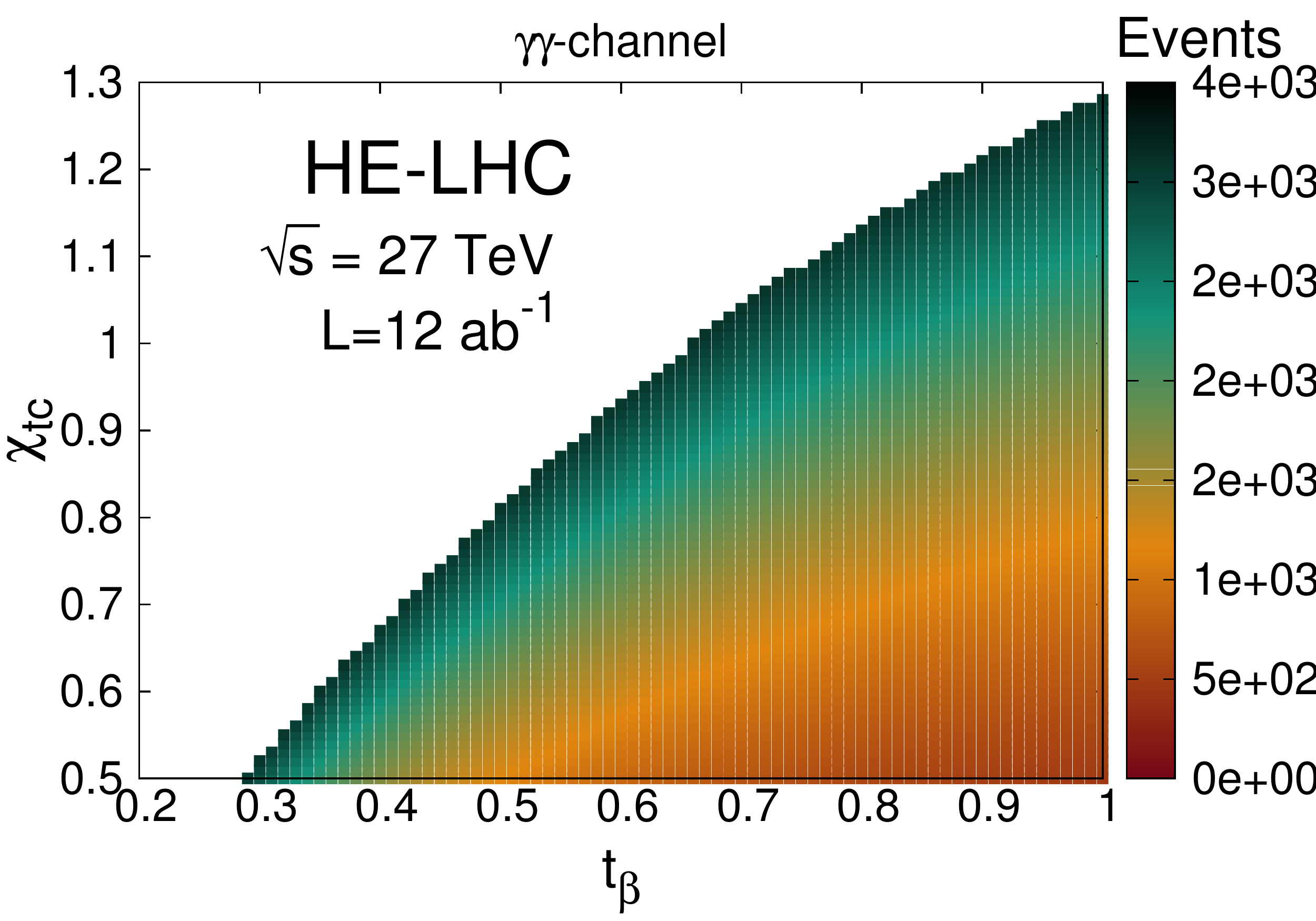}}
\subfigure[]{\includegraphics[scale = 0.25,angle=270]{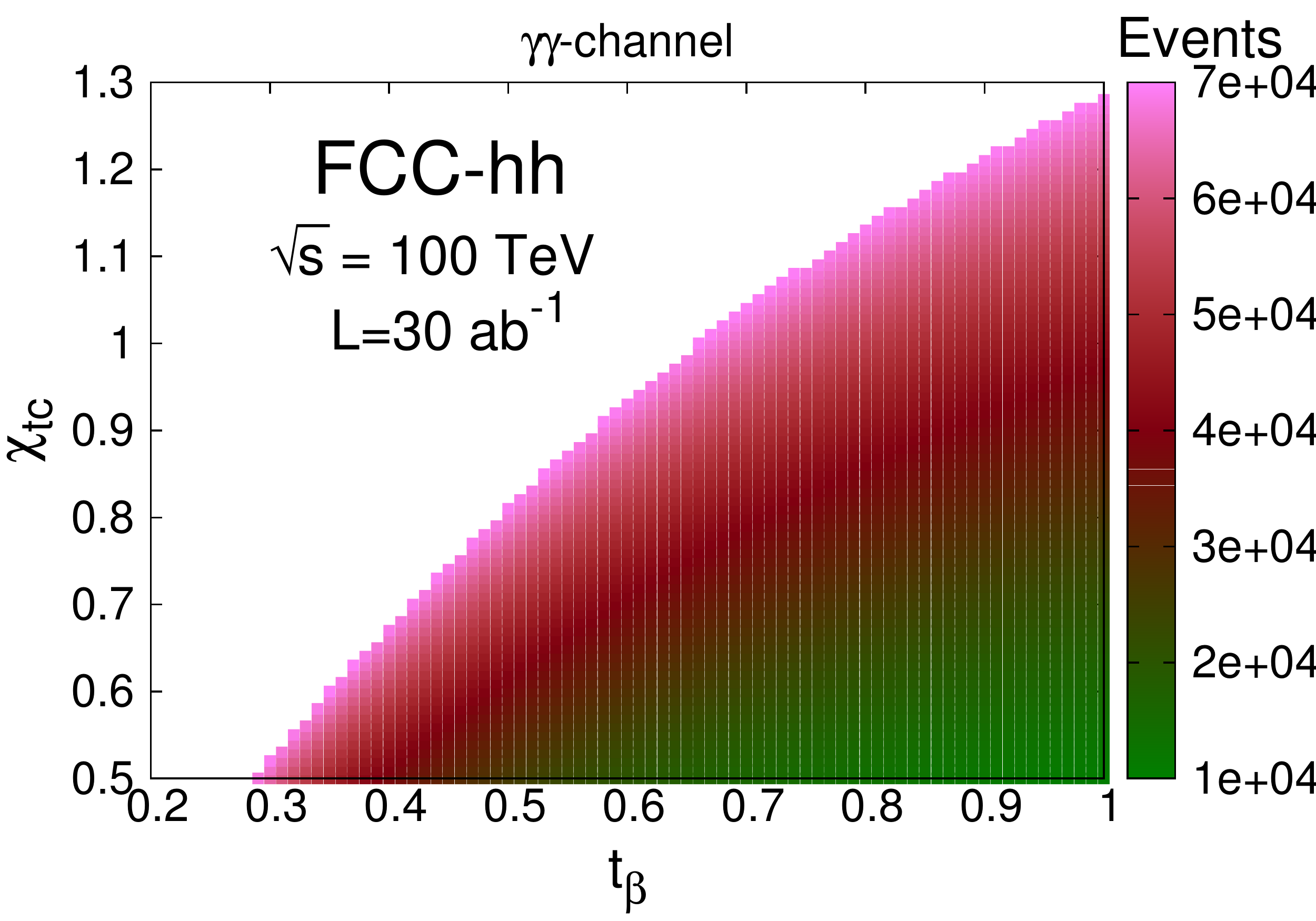}}
 \caption{Number of signal events for the \textit{$\gamma\gamma$-channel} in the $t_{\beta}-\chi_{tc}$ plane: (a) LHC to $\mathcal{L}$=0.3 ab$^{-1}$, (b) HL-LHC to $\mathcal{L}$=3 ab$^{-1}$, (c) HE-LHC to $\mathcal{L}$=12 ab$^{-1}$ and (d) FCC-hh $\mathcal{L}$=30 ab$^{-1}$. \label{EventsGAGAChannel}}
	\end{figure}

The figure \ref{EventsGAGAChannel} shows the number of signal events produced at the LHC, HL-LHC, HE-LHC and FCC-hh with  integrated luminosities of 0.3 ab$^{-1}$, 3 ab$^{-1}$, 12 ab$^{-1}$ and 30 ab$^{-1}$ and center-of-mass energies of $\sqrt{s}=14$ TeV (LHC), $\sqrt{s}=14$ TeV (HL-LHC), $\sqrt{s}=27$ TeV and $\sqrt{s}=100$ TeV, respectively.
In all cases (a)-(d), the number of events is high when $t_{\beta}$ as increase as $\chi_{tc}$, which is expected since the $htc$ coupling behaves as $\sim \chi_{tc}/t_{\beta}$. For the benchmark points $(t_{\beta}\sim 0.3,\,\chi_{tc}\sim 0.5)$ and $(t_{\beta}\sim 1,\,\chi_{tc}\sim 1.3)$, the number of signal events are 30, 300, $4\times 10^3$ and $7\times 10^4$ for LHC, HL-LHC, HE-LHC and FCC-hh, respectively. If $t_{\beta}$ is fixed and $\chi_{tc}$ is scanned, the number of signal events increase. Otherwise, if $\chi_{tc}$ is fixed and $t_{\beta}$ is scanned, the number of signal event decreases.


\subsubsection{$bb$ channel}
\begin{figure}[!h]
\centering
\subfigure[]{\includegraphics[scale = 0.25,angle=270]{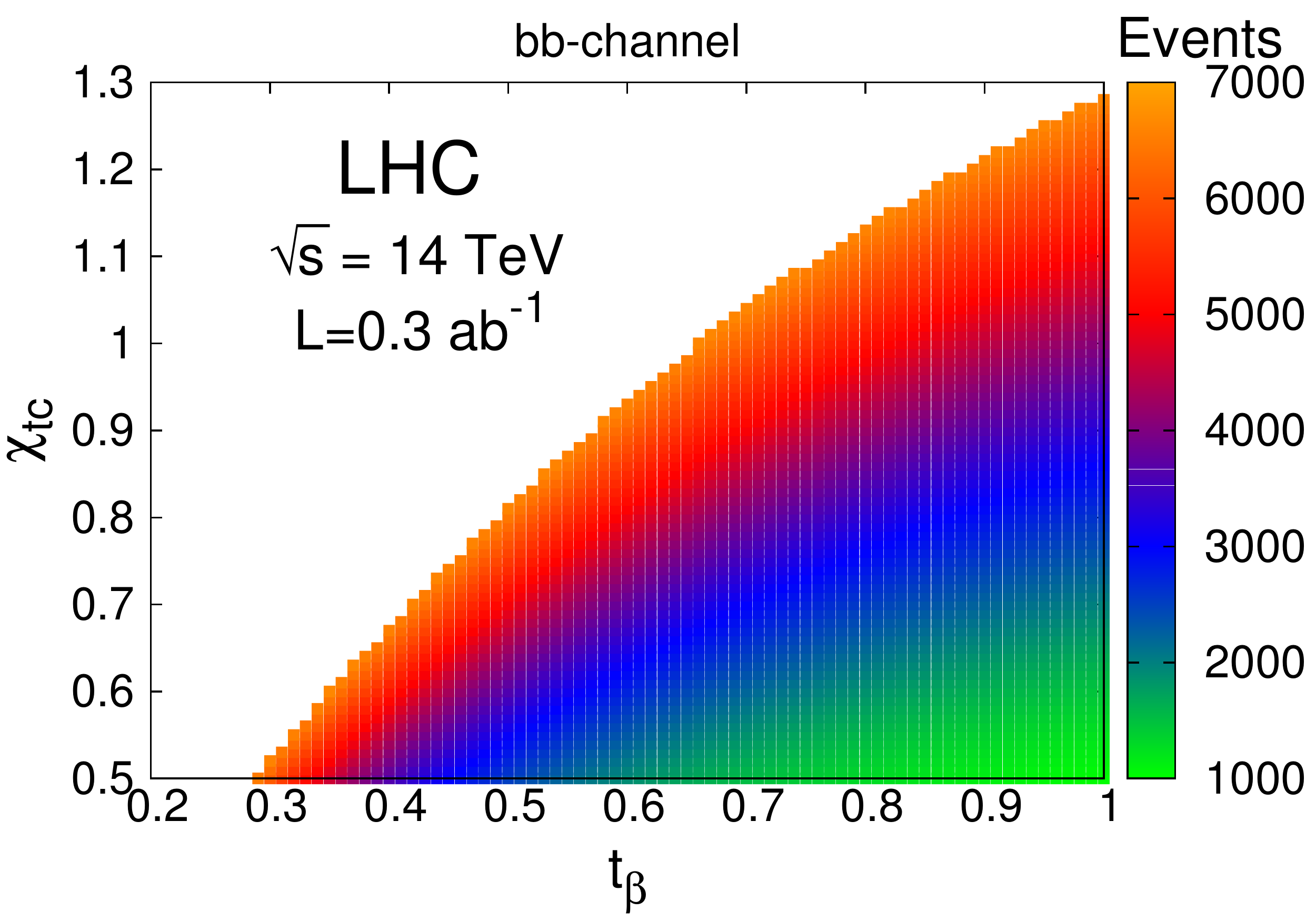}}
\subfigure[]{\includegraphics[scale = 0.25,angle=270]{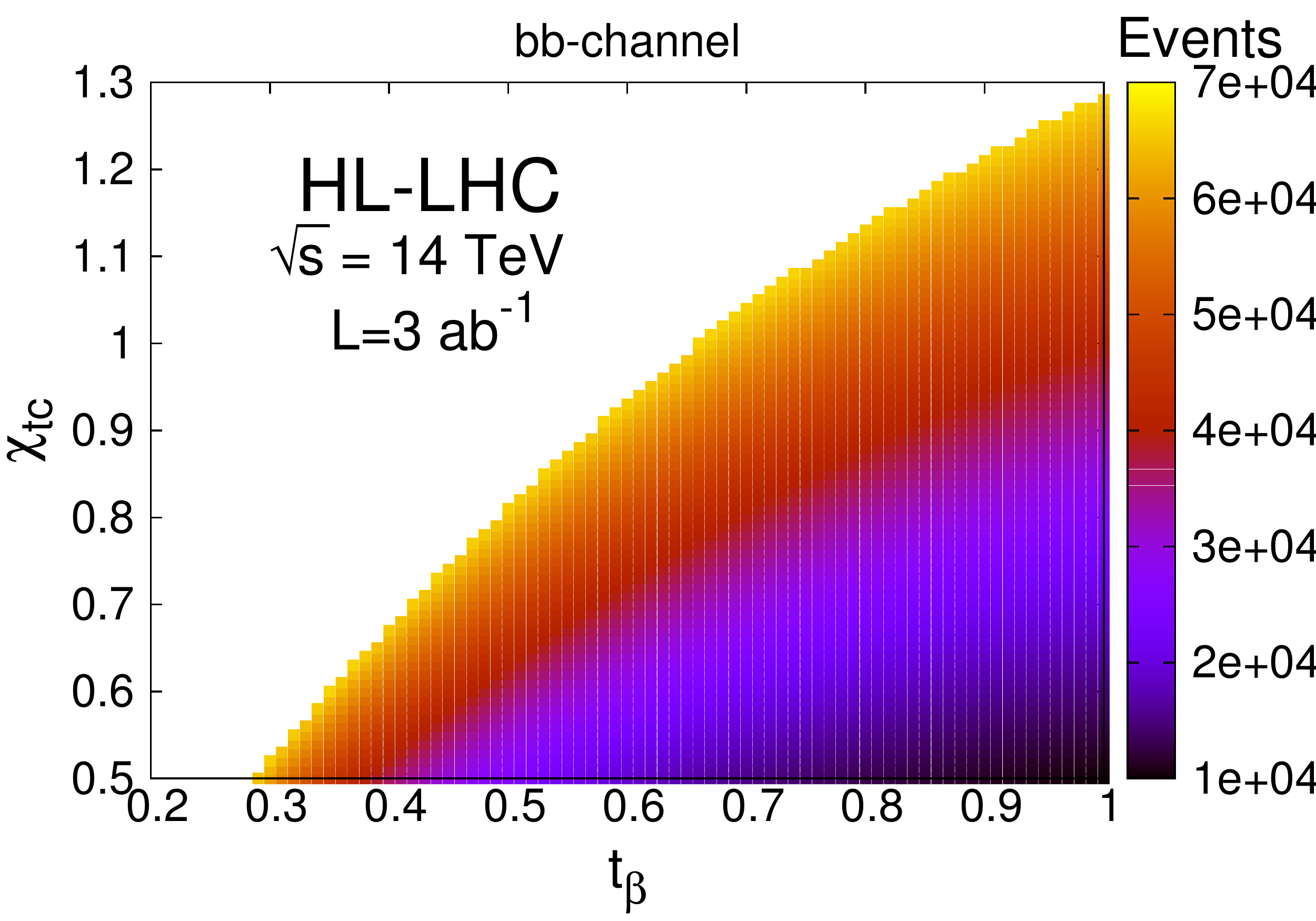}}
\subfigure[]{\includegraphics[scale = 0.25,angle=270]{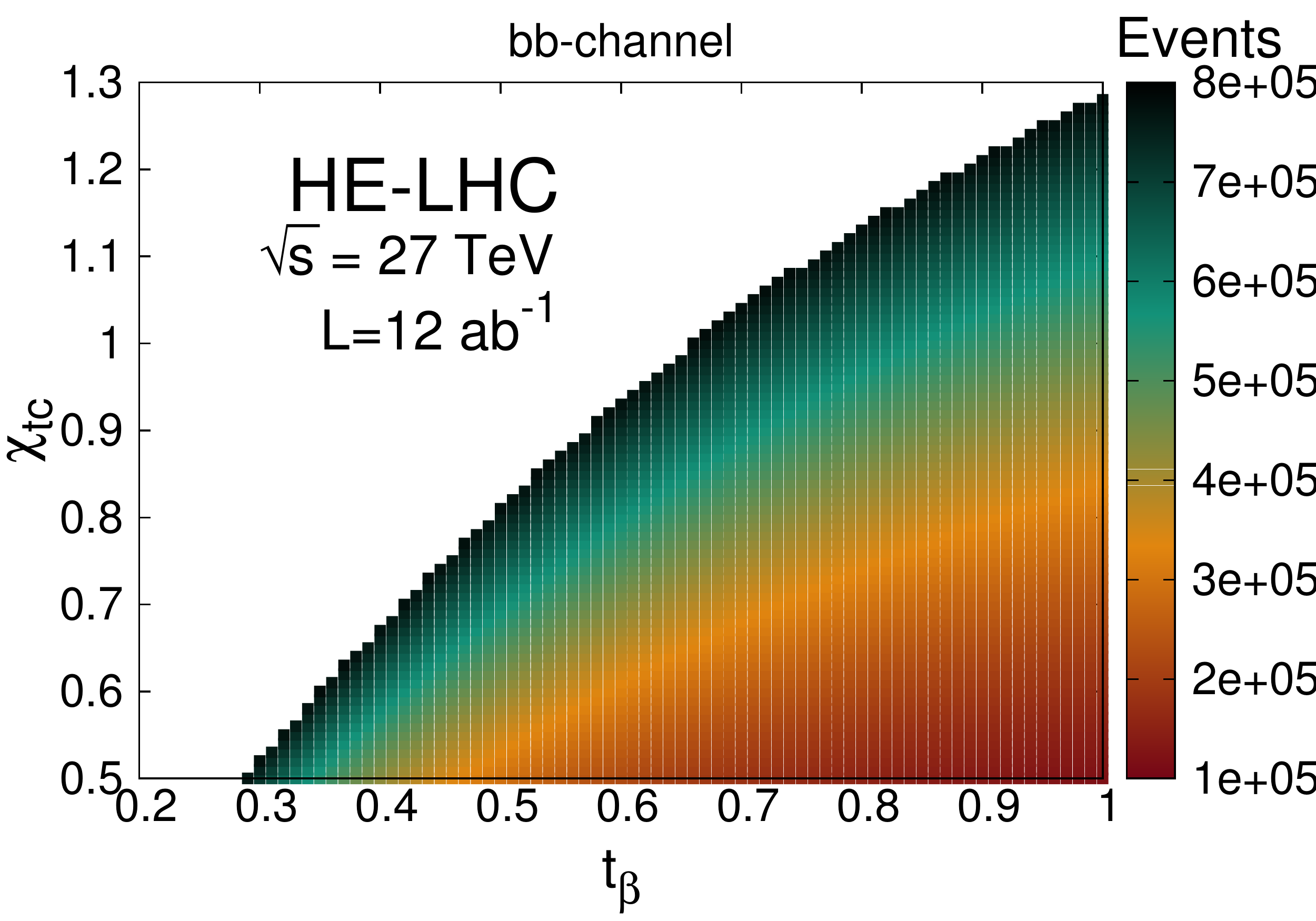}}
\subfigure[]{\includegraphics[scale = 0.25,angle=270]{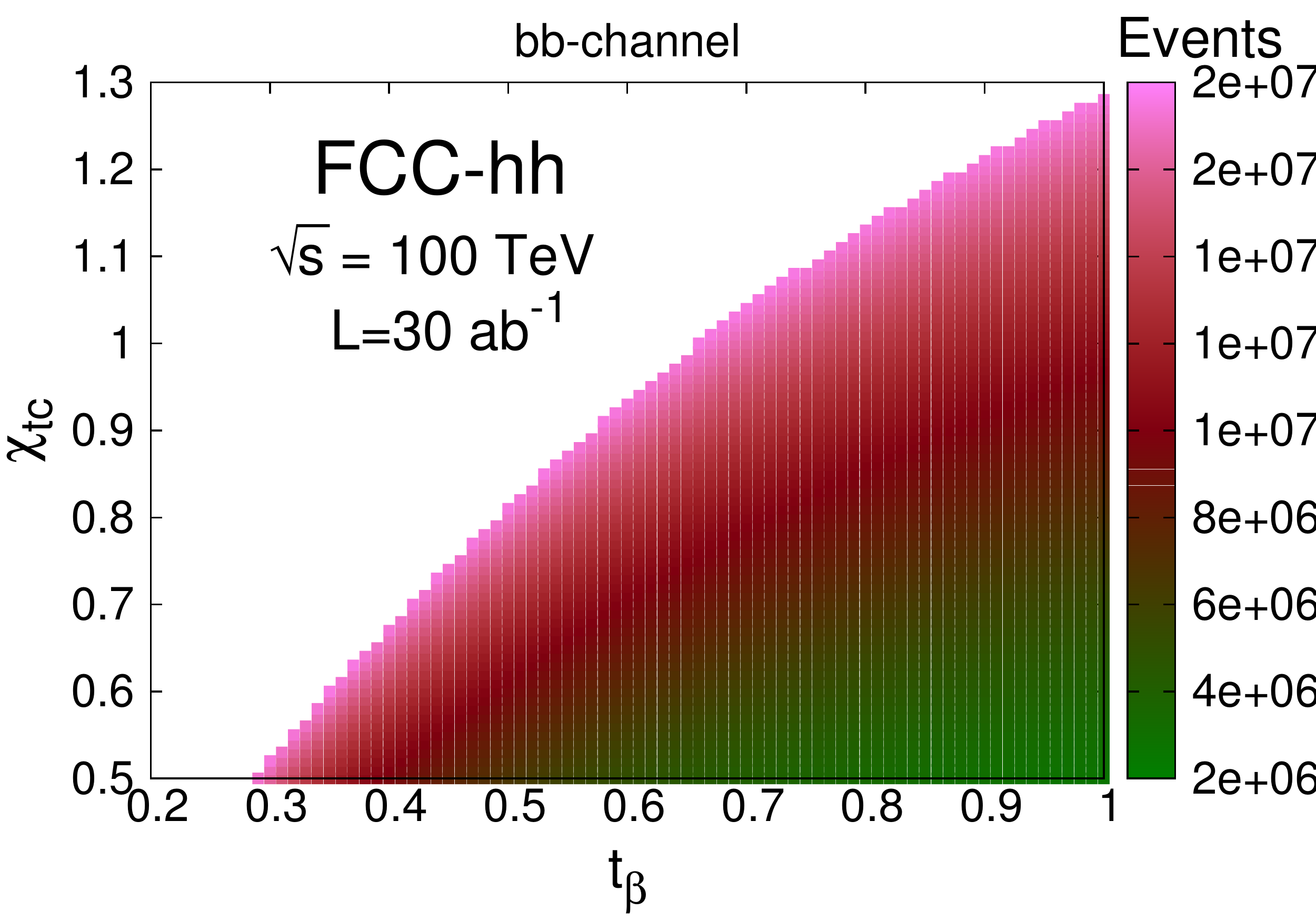}}
 \caption{Number of signal events for the \textit{$bb$-channel} in the $t_{\beta}-\chi_{tc}$ plane: (a) LHC to $\mathcal{L}$=0.3 ab$^{-1}$, (b) HL-LHC to $\mathcal{L}$=3 ab$^{-1}$, (c) HE-LHC to $\mathcal{L}$=12 ab$^{-1}$ and (d) FCC-hh $\mathcal{L}$=30 ab$^{-1}$. \label{EventsBBChannel}}
	\end{figure}

As far as \textit{$bb$-channel} is concerned, the figure \ref{EventsBBChannel} presents the same as in figure \ref{EventsGAGAChannel} though for the \textit{$bb$-channel}.
As the $\textit{$\gamma\gamma$-channel}$, the number of signal events of the $\textit{bb-channel}$ behave very similar. However, because the $\mathcal{B}(h\to b\bar{b})\sim 10^2\cdot\mathcal{B}(h\to\gamma\gamma)$, the number of signal events increase about two orders of magnitude being $7\times 10^3$, $7\times 10^4$, $8\times 10^5$, $2\times 10^7$ for LHC, HL-LHC, HE-LHC and FCC-hh, respectively. The $\textit{bb-channel}$ gives a great opportunity to detect the signal, as discussed below.	
	\subsection{Signal and SM dominant background simulation}
	Signal events are produced through $t\bar{t}$ production at the hadron colliders considered, the first top decays into a Higgs boson and a charm quark and, the second one, into a bottom quark, a light charged lepton plus a neutrino via a $W$ gauge boson. In the $\textit{$\gamma\gamma$-channel}$ the Higgs boson decays into two photons and in the $bb$-channel the Higgs boson decays into two bottom quarks. As far as the computation scheme is concerned, the Feynman rules in the THDM-III were implemented via $\texttt{LanHEP}$ routines \cite{Semenov:2014rea} for a \texttt{UFO} model \cite{Degrande:2011ua}. $10^5$ parton-level events were generated for the signal and the SM main background using $\texttt{MadGraph5}$ \cite{Alwall:2011uj} and perform shower and hadronization with $\texttt{Pythia8}$ \cite{Sjostrand:2006za}. The $\texttt{CT10}$ parton distribution function \cite{Gao:2013xoa} is used. A Higgs boson mass of 125 GeV and a top quark mass of 173 GeV were considered \cite{Tanabashi:2018oca}. Afterwards, the kinematic analysis was done via $\texttt{MadAnalysis5}$ \cite{Conte:2012fm}. As far as the jet reconstruction, the jet finding package $\texttt{FastJet}$ \cite{Cacciari:2011ma} and the anti$-k_T$ algorithm, with $R=0.4$, were used, which are implemented in $\texttt{MadAnalysis5}$.
\subsubsection{Mass reconstruction}
\subsubsection*{$\textit{$\gamma\gamma$-channel}$}
\begin{figure}[!h]
\centering
\subfigure[]{\includegraphics[scale = 0.26]{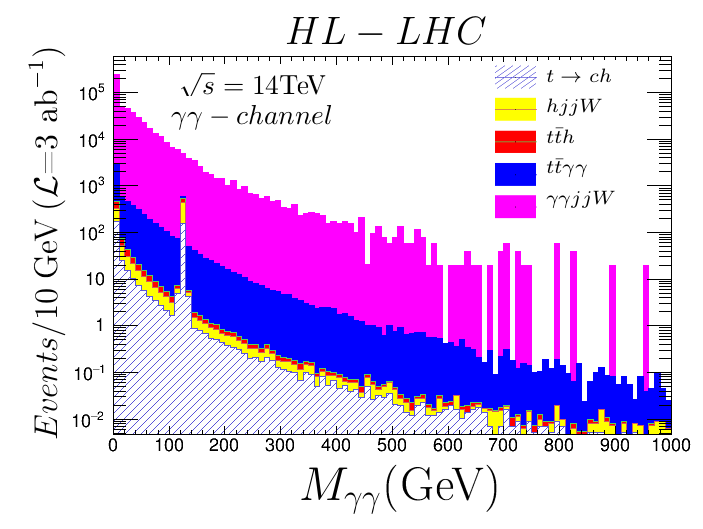}}
\subfigure[]{\includegraphics[scale = 0.26]{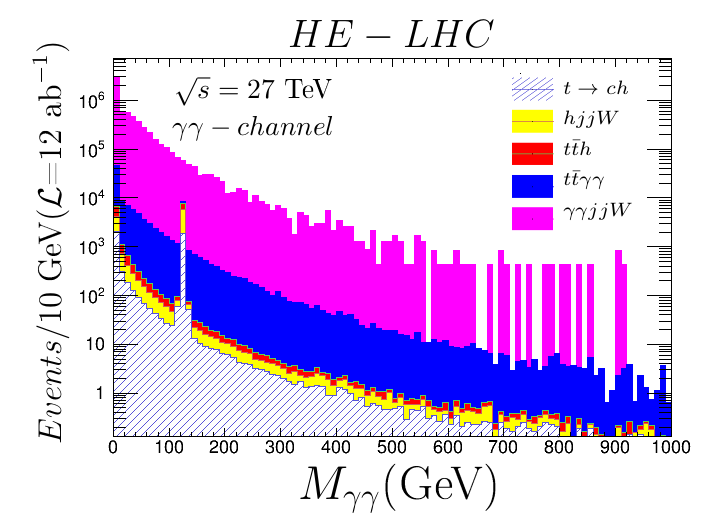}}
\subfigure[]{\includegraphics[scale = 0.26]{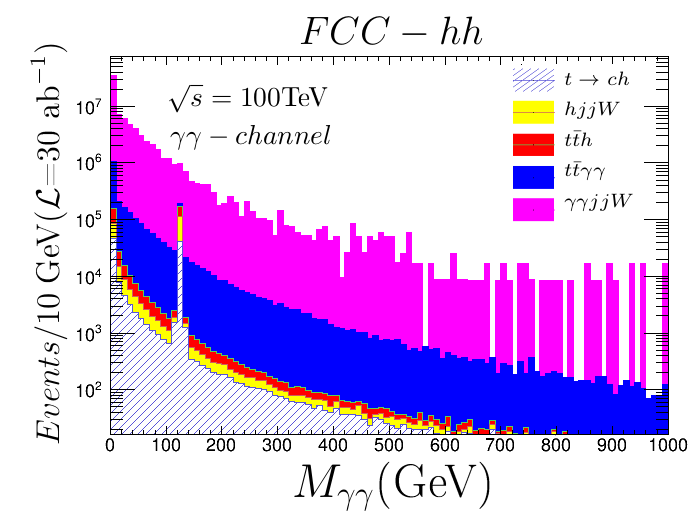}}
 \caption{Invariant mass distribution of the diphoton system, $M_{\gamma\gamma}$, without cuts.
 \label{InvMass_GamGam_ALL}}
\end{figure}

Since the signal comes from $(pp\to t\bar{t},\,t\to ch,\,h\to\gamma\gamma)$, the Higgs boson mass was reconstructed as the invariant mass of the diphoton system, $M_{\gamma\gamma}$. Events which the invariant mass is between $123-127$ GeV were selected, as we discussed below.
The figure \ref{InvMass_GamGam_ALL} shows the invariant mass distribution $M_{\gamma\gamma}$ without cuts.

\subsubsection*{$\textit{bb-channel}$}
\begin{figure}[!h]
\centering
\subfigure[]{\includegraphics[scale = 0.26]{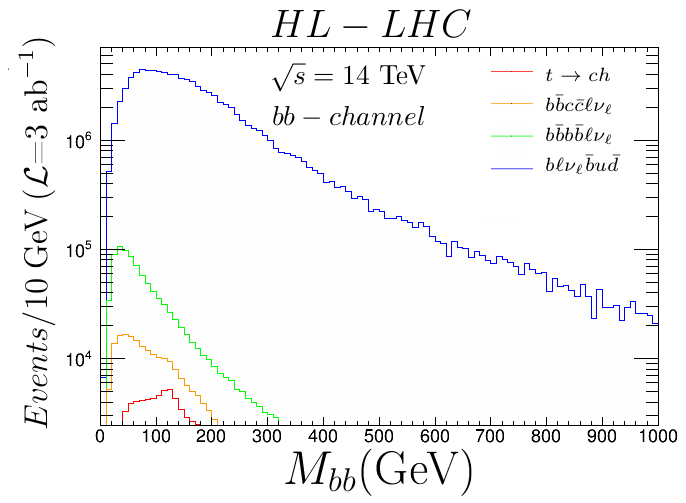}}
\subfigure[]{\includegraphics[scale = 0.26]{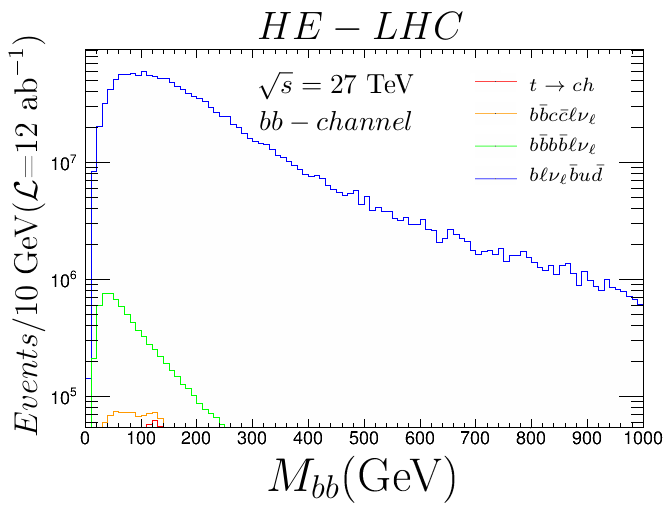}}
\subfigure[]{\includegraphics[scale = 0.26]{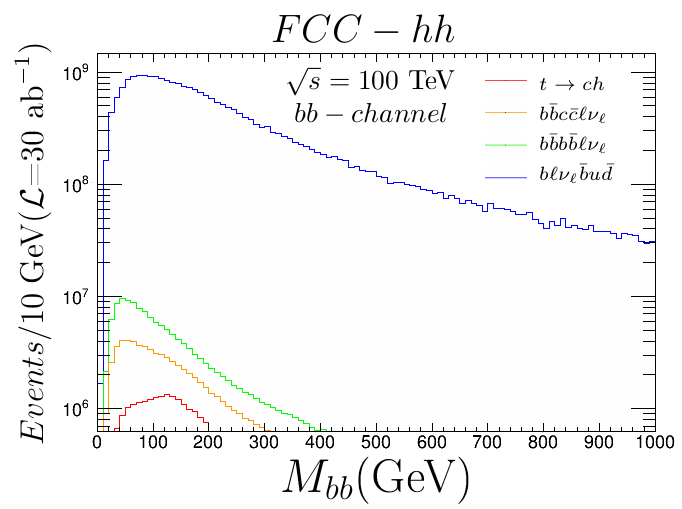}}
 \caption{Invariant mass distribution of the $bb$ system, $M_{bb}$, without cuts. \label{InvMass_BB_ALL}}
\end{figure}

In this channel, the signal comes from $(pp\to t\bar{t},\,t\to ch,\,h\to b\bar{b})$, as for the $\textit{$\gamma\gamma$-channel}$, the Higgs boson mass was reconstructed as the invariant mass, but now for a $bb$ pair, such that $|M_{b_1 b_2}-m_h|\leq 0.15m_h$. The figure \ref{InvMass_BB_ALL} presents the invariant mass distribution, $M_{bb}$, without cuts.

\subsubsection{Kinematic cuts}
In order to isolate the signal, the following kinematic cuts were applied.
\subsubsection*{$\textit{$\gamma\gamma$-channel}$\label{cutsgamgam}}
For both signal and background events the following kinematic cuts were imposed:
\begin{itemize}
\item Exactly one $b-$jet and two photons.
\item We identify leptons and photons by imposing:  $p_T^{\gamma,\,\ell}>$ 25 GeV.
\item The invariant mass of the diphoton system, $M_{\gamma\gamma}$, is the main variable for search the Higgs boson decay, events between: $123\leq M_{\gamma\gamma}\leq 127$ GeV are acepted.
\item Because the Higgs boson decays into two photons, in order to reconstruct the signal top quark from the identified $b-$jet and the diphoton system, it is required that: $160\leq M_{\gamma\gamma j}\leq 190$ GeV.
\item The distance between photons coming from Higgs boson decay and the distance between the diphoton system and the jet must be: 1.8 $<\Delta R(\gamma,\,\gamma)<$ 5.0, $\Delta R(\gamma\gamma,\,j)<$ 1.8.
\item The ATLAS collaboration reported in the ref. \cite{btagATLAS}, that the $b$ tagging efficiency ($\epsilon_b$) is $\sim70\%$, the probability that a $c$-jet is mistagged as a $b$-jet ($\epsilon_c$) is of the order of $10\%$ \cite{LHCbcjet}, while the probability that any other jet is mistagged as a $b$-jet ($\epsilon_j$) is of the order of $1\%$.
Following it, the tagging and mistagging efficiencies considered in this work are as follows:
\begin{itemize}
\item $\epsilon_b=70\%$,
\item $\epsilon_c=14\%$,
\item $\epsilon_j=1\%$.
\end{itemize}
\end{itemize}

\subsubsection*{$\textit{bb-channel}$\label{cuts}}	
For both signal and background events there should be:
\begin{itemize}
\item Exactly four jets: three of them are tagged as $b-$jets with $p_T^{j,\,b}>$30 GeV and $|\eta^j|<$ 2.5.
\item Exactly one isolated lepton with: $p_T^{\ell}>$20 GeV, $|\eta^{\ell}|<$2.5.
\item Because in the final state emerge a neutrino, the missing transverse energy (MET) must be MET$>$ 30 GeV.
\item In order to reconstruct the top quark mass associated with the FCNC, it is required that $|M_{b_1 b_2 j}-m_t|\leq$ 26 GeV.
\item In order to reconstruct the Higgs boson mass as the invariant mass of the $bb$ system, it is imposed that:  $|M_{b_1b_2}-m_h|\leq$ 0.15$m_h$.
\item It is required that $\Delta R$ is between each jet and that charged lepton pair is $\sqrt{\Delta\phi^2+\Delta\eta^2}>$ 0.4
\item The tagging and mistagging efficiencies are as follows
\begin{itemize}
\item $\epsilon_b=70\%$,
\item $\epsilon_c=14\%$,
\item $\epsilon_j=1\%$.
\end{itemize}
\end{itemize}

\subsection{Evidence and potential discovery}
In this section we compute the signal significance defined as $\mathcal{S}=N_S/\sqrt{N_S+N_B}$, where $N_S$ are the number of signal events and $N_B$ is the number of SM background events once the kinematic cuts were applied.
\subsubsection{$\textit{$\gamma\gamma$-channel}$}
\begin{figure}[!h]
\centering
\subfigure[]{\includegraphics[scale = 0.24,angle=270]{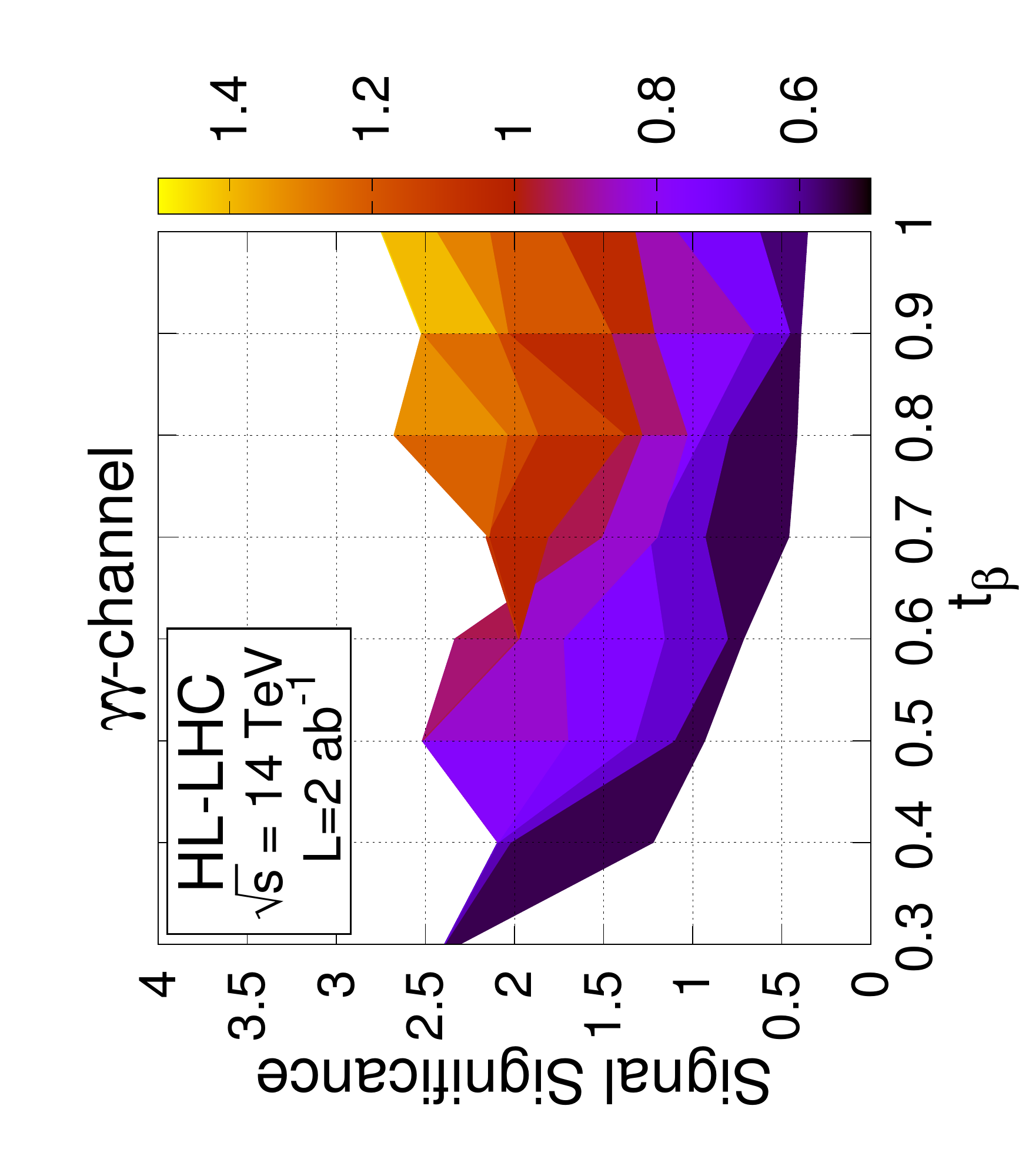}}
\subfigure[]{\includegraphics[scale = 0.24,angle=270]{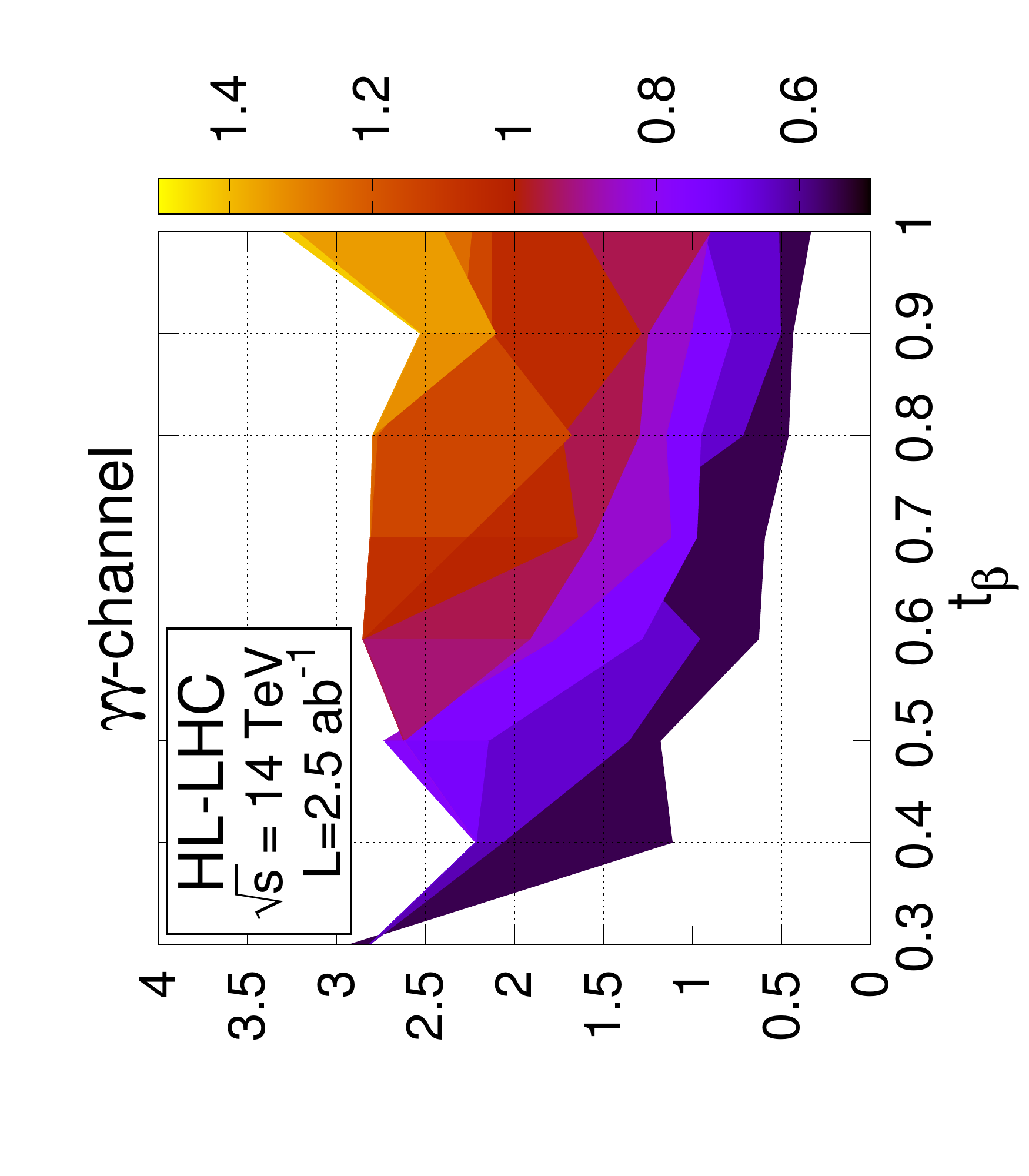}}
\subfigure[]{\includegraphics[scale = 0.24,angle=270]{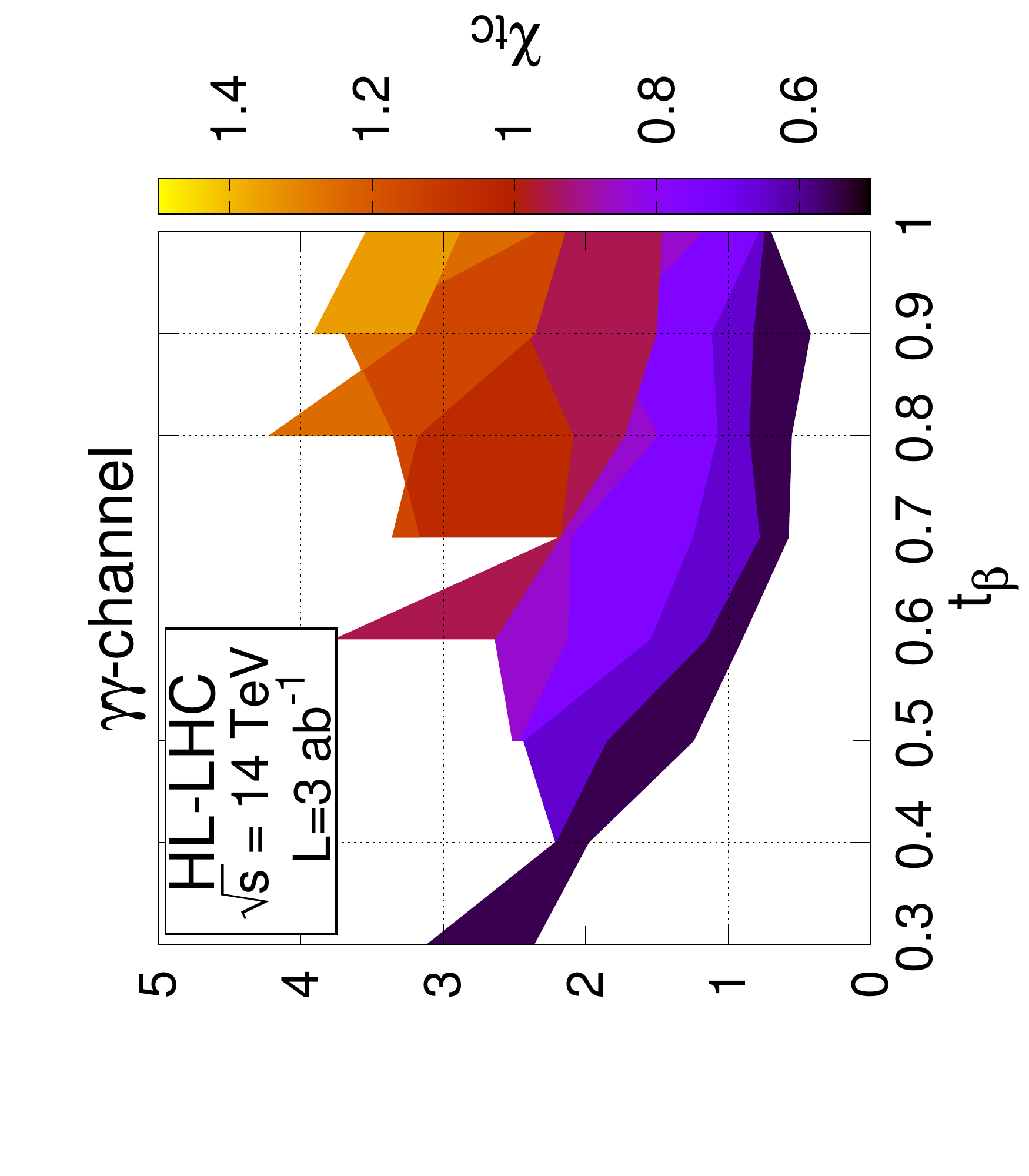}}
 \caption{Density plots for the signal significance as a function of the $t_{\beta}$ and $\chi_{tc}$ for three illustrative integrated luminosities: (a) $\mathcal{L}$=2 ab$^{-1}$, (b) $\mathcal{L}$=2.5 ab$^{-1}$, (c) $\mathcal{L}$=3 ab$^{-1}$. The case (c) represents the final integrated luminosity reached by the HL-LHC. \label{SigmaGamGamChannel_HL-LHC}}
\end{figure}

\begin{figure}[!h]
\centering
\subfigure[]{\includegraphics[scale = 0.24,angle=270]{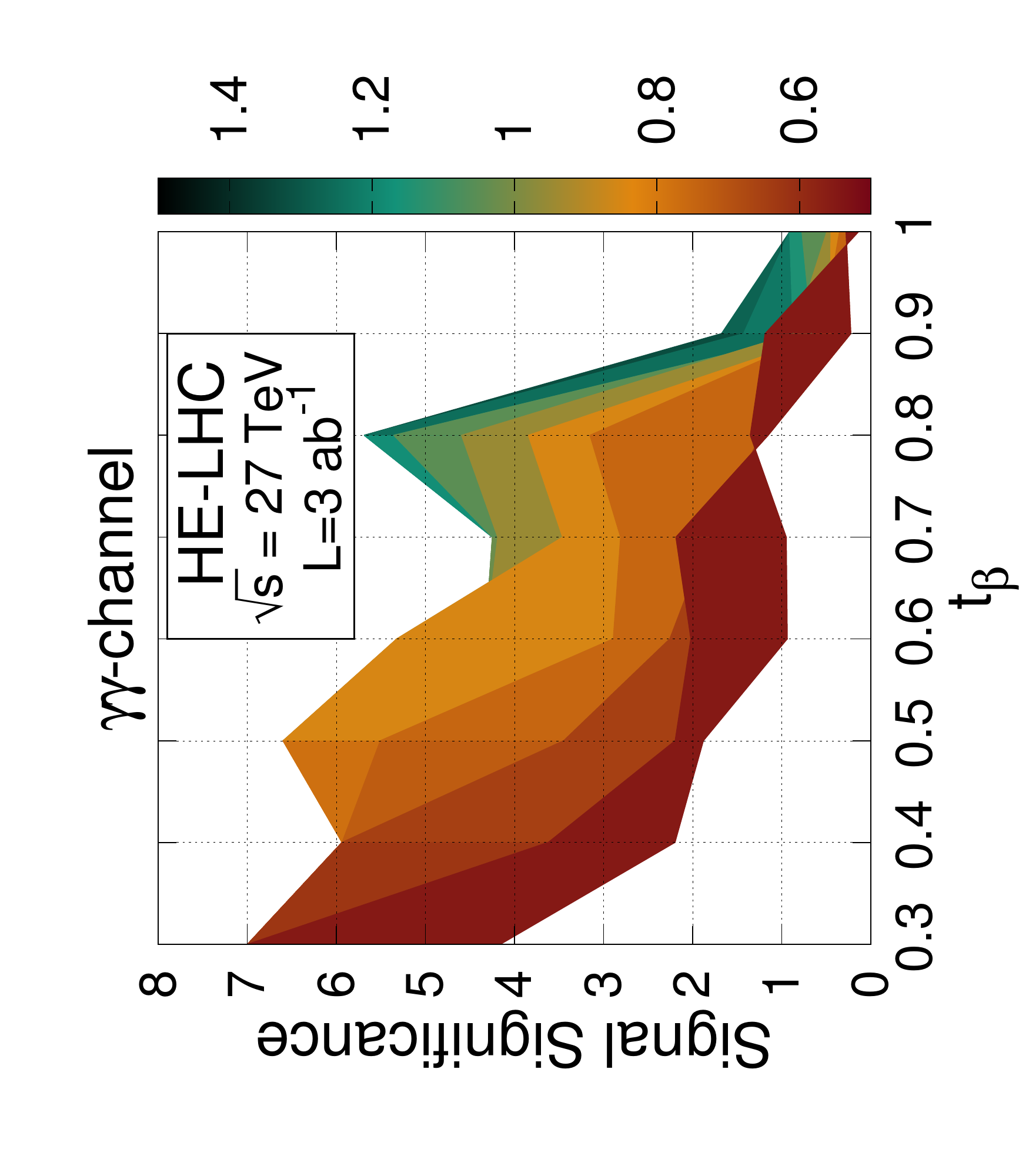}}
\subfigure[]{\includegraphics[scale = 0.24,angle=270]{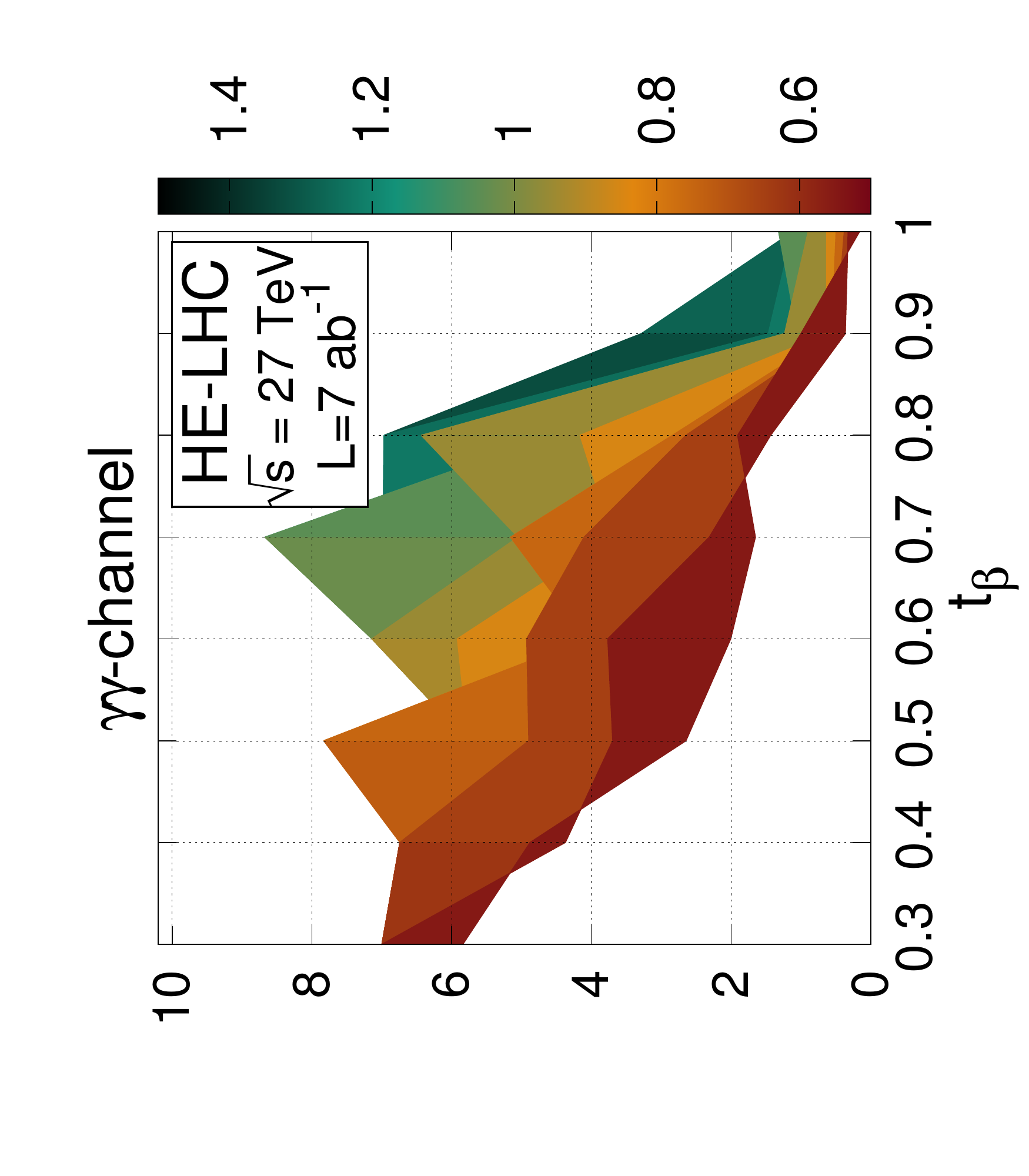}}
\subfigure[]{\includegraphics[scale = 0.24,angle=270]{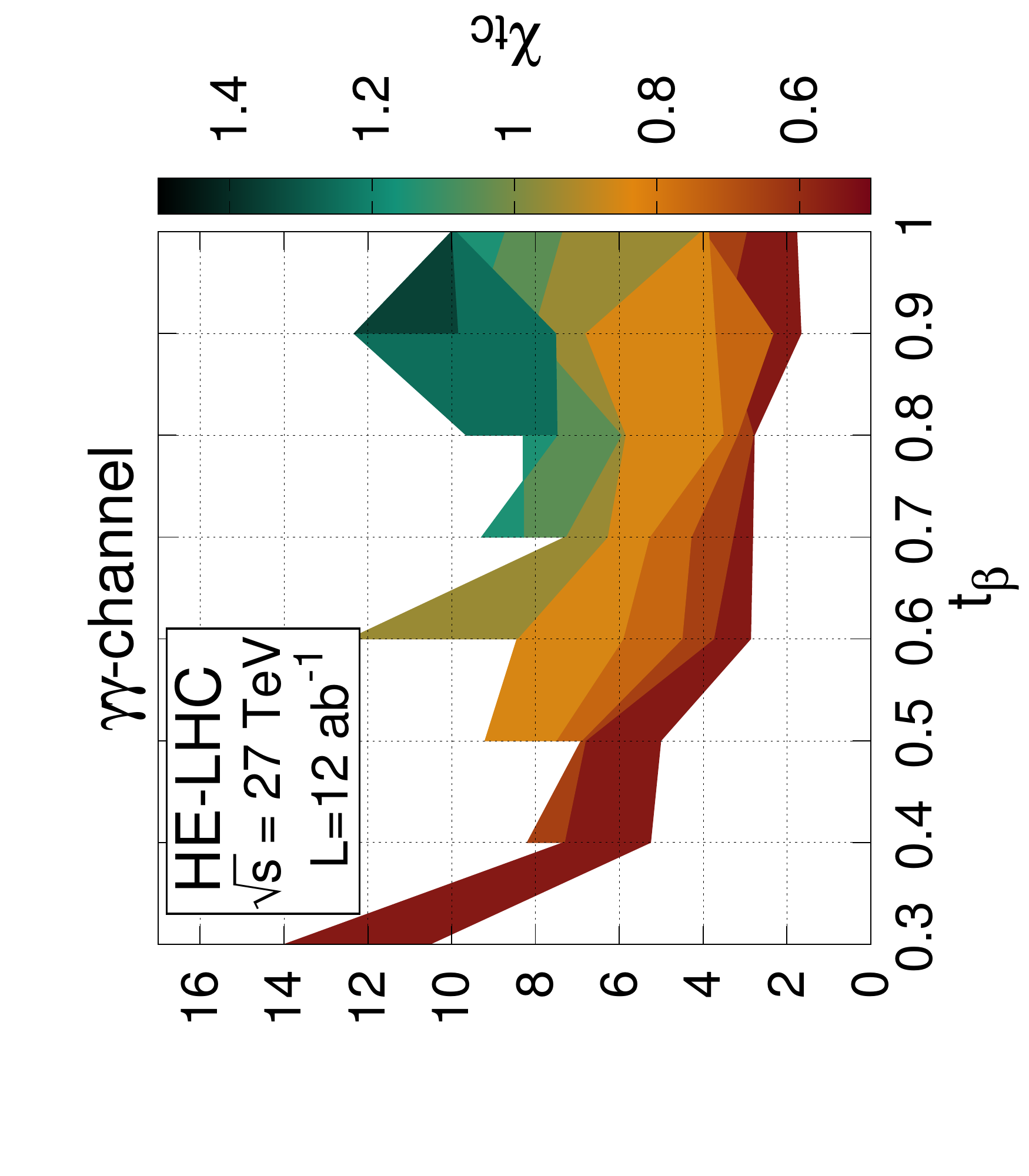}}
 \caption{The same as in figure \ref{SigmaGamGamChannel_HL-LHC} but for: (a) $\mathcal{L}$=3 ab$^{-1}$, (b) $\mathcal{L}$=7 ab$^{-1}$, (c) $\mathcal{L}$=12 ab$^{-1}$. The case (c) represents the final integrated luminosity reached by the HE-LHC. \label{SigmaGamGamChannel_HE-LHC}}
\end{figure}

\begin{figure}[!h]
\centering
\subfigure[]{\includegraphics[scale = 0.24,angle=270]{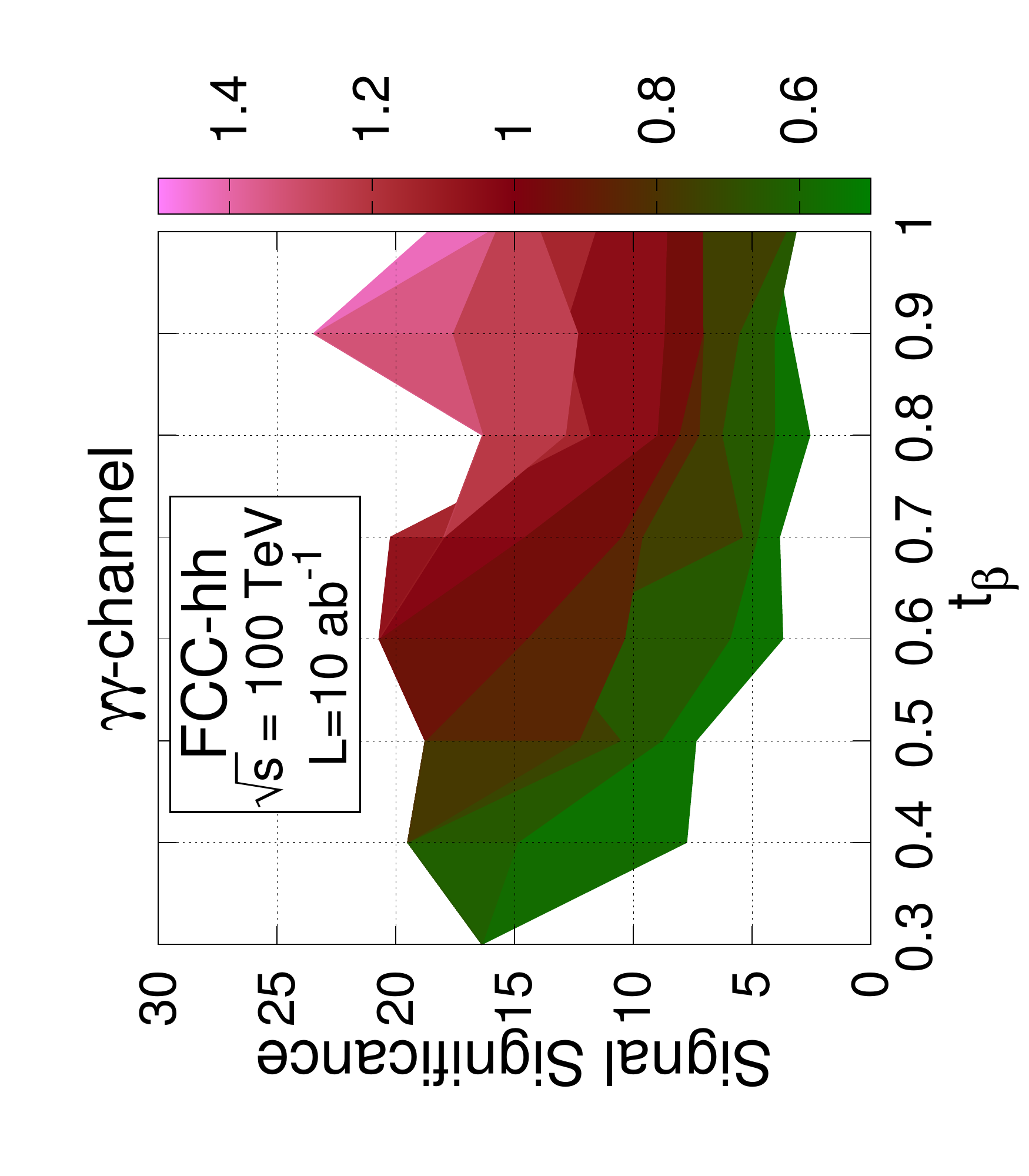}}
\subfigure[]{\includegraphics[scale = 0.24,angle=270]{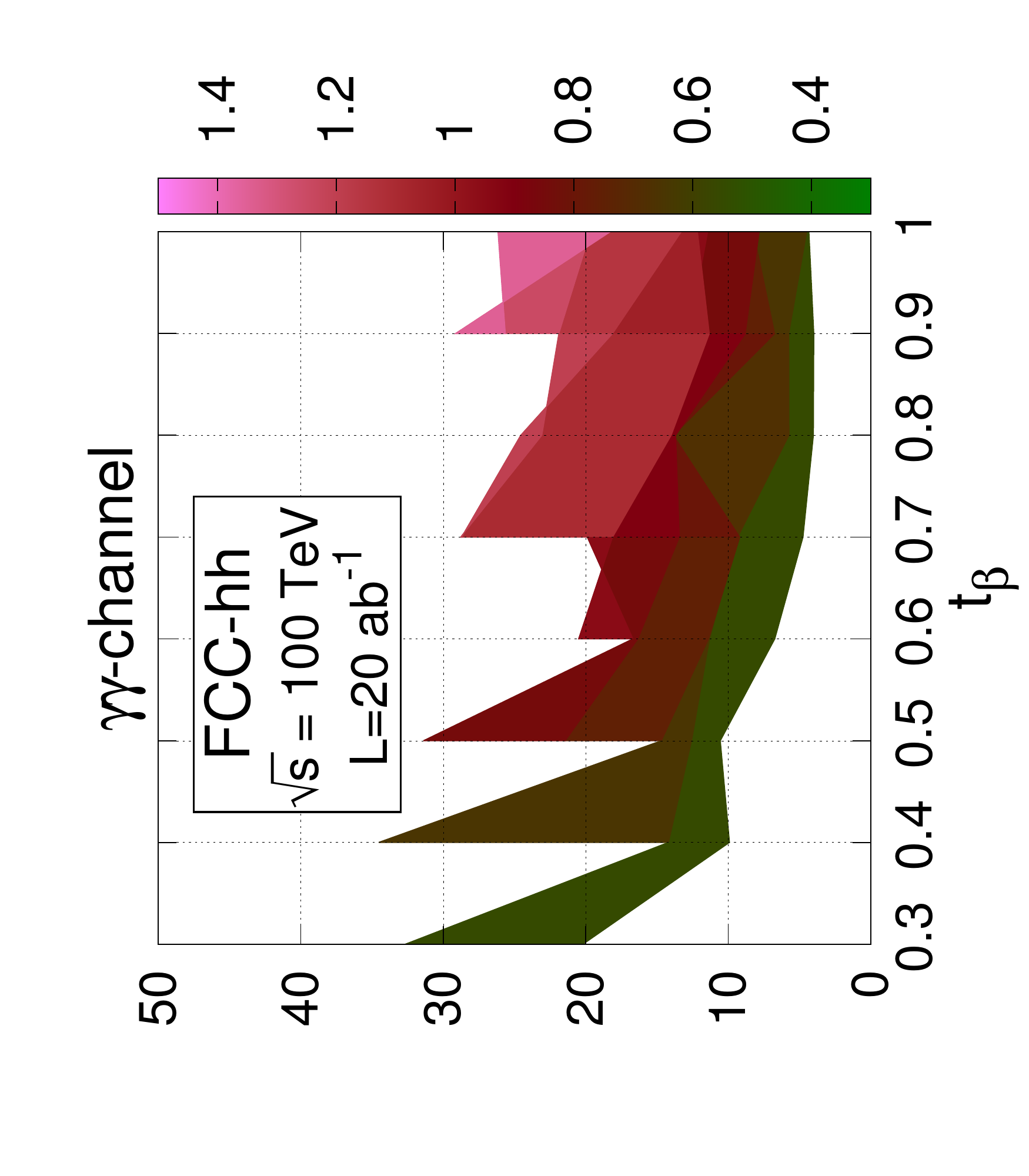}}
\subfigure[]{\includegraphics[scale = 0.24,angle=270]{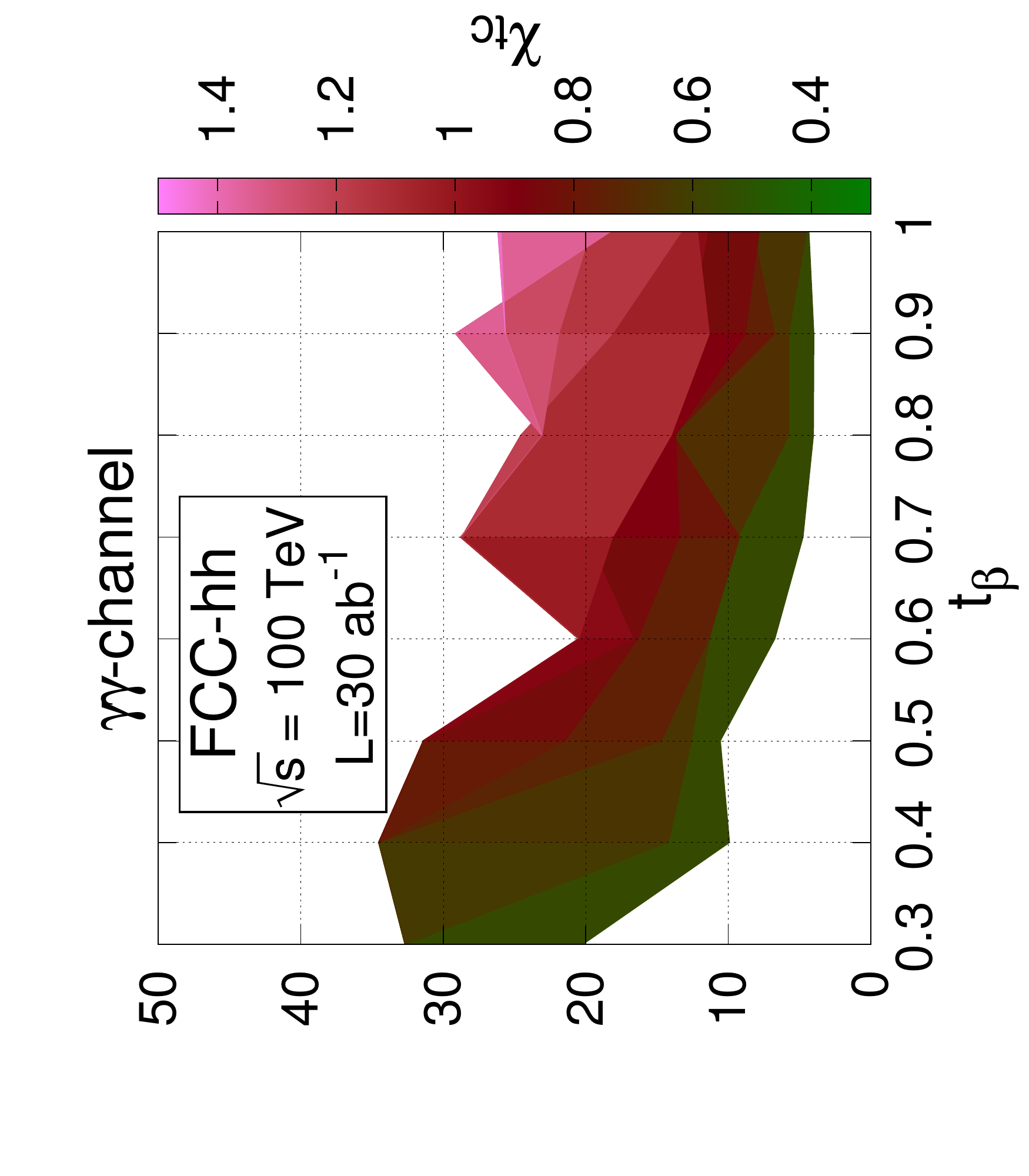}}
 \caption{The same as in figure \ref{SigmaGamGamChannel_HL-LHC} but for: (a) $\mathcal{L}$=10 ab$^{-1}$, (b) $\mathcal{L}$=20 ab$^{-1}$, (c) $\mathcal{L}$=30 ab$^{-1}$. The case (c) represents the final integrated luminosity reached by the FCC-hh. \label{SigmaGamGamChannel_FCC-hh}}
\end{figure}

After applying the kinematic cuts shown in section \ref{cutsgamgam}, evidence for the $t\to ch$ decay in the $\textit{$\gamma\gamma$-channel}$ with a integrated luminosity of $\sim 3$ ab$^{-1}$ is found. Density plots of the signal significance as a function of $t_{\beta}$ and $\chi_{tc}$ are presented in the figure \ref{SigmaGamGamChannel_HL-LHC}. Three illustrative integrated luminosities which will be achieved at the HL-LHC, namely, $\mathcal{L}$=2, 2.5, 3 ab$^{-1}$ are considered. It is found a region between $0.6\leq t_{\beta}\leq 1$ and $0.9\leq\chi_{tc}\leq 1.3$ intervals, with a signal significance $3\sigma\leq\mathcal{S}$, which allows us to claim evidence for $t\to ch$ decay. The figure \ref{SigmaGamGamChannel_HE-LHC} shows the same as in the figure \ref{SigmaGamGamChannel_HL-LHC} but for the HE-LHC. We found that with an integrated luminosity of $\sim0.3$ ab$^{-1}$ ($300$ fb$^{-1}$), evidence for the $t\to hc$ decay would be established. However, higher standard deviations may be achieved which range from $7\sigma$ ($\mathcal{L}$=3 ab$^{-1}$) to $14\sigma$ ($\mathcal{L}$=12 ab$^{-1}$). This collider could be used, among other things, to perform several cross-checks of the discovery of $t\to ch$ decay. Finally, the figure \ref{SigmaGamGamChannel_FCC-hh} presents density plots for the FCC-hh collider. Signal significances of the order of $\mathcal{O}(30)$ are found. This means, along with $\textit{bb-channel}$, as we will discuss below, an opportunity to secure new physics and focus on finding new sources of physics beyond the SM.

\subsubsection{$\textit{bb-channel}$}
\begin{figure}[!h]
\centering
\subfigure[]{\includegraphics[scale = 0.24,angle=270]{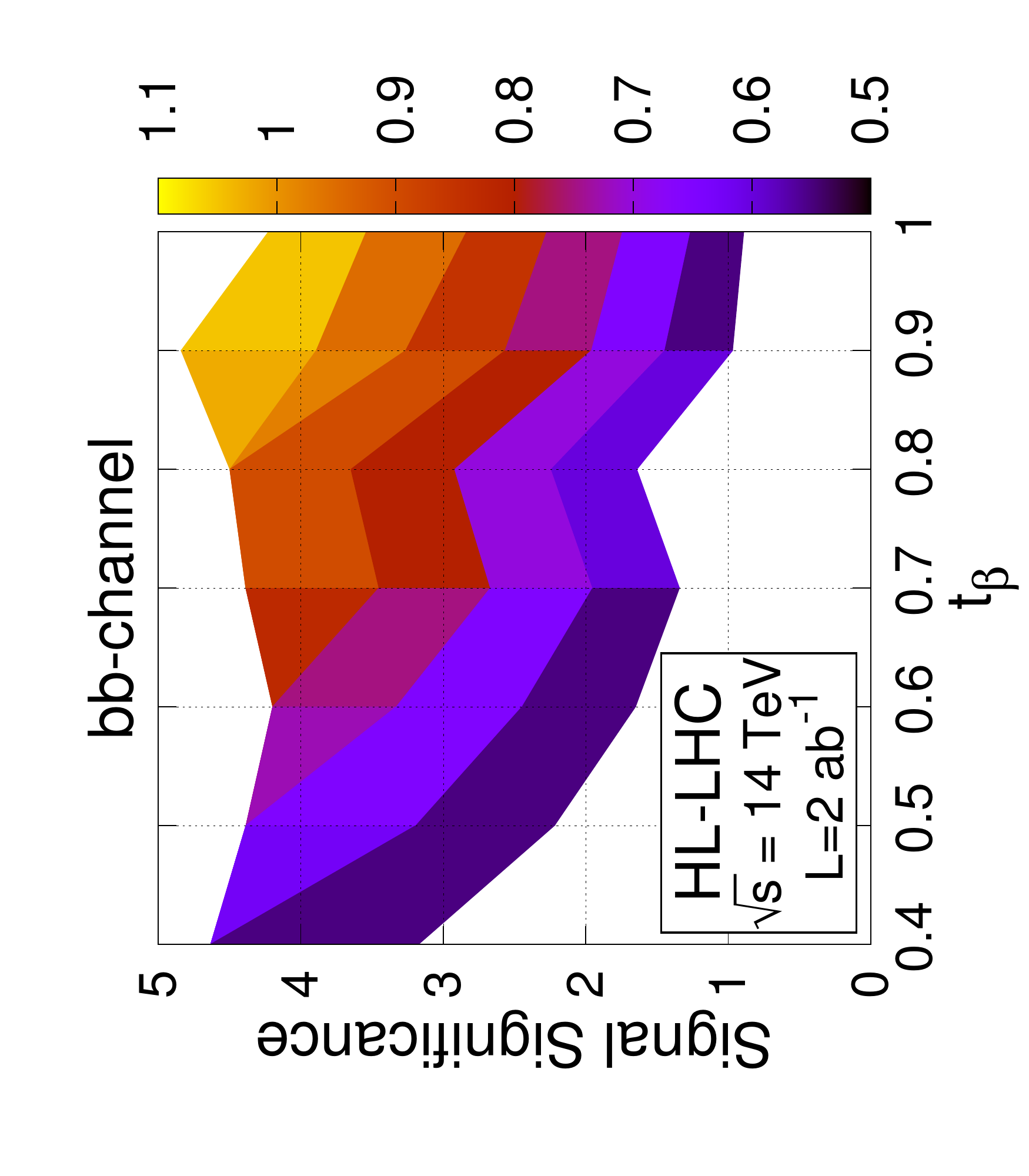}}
\subfigure[]{\includegraphics[scale = 0.24,angle=270]{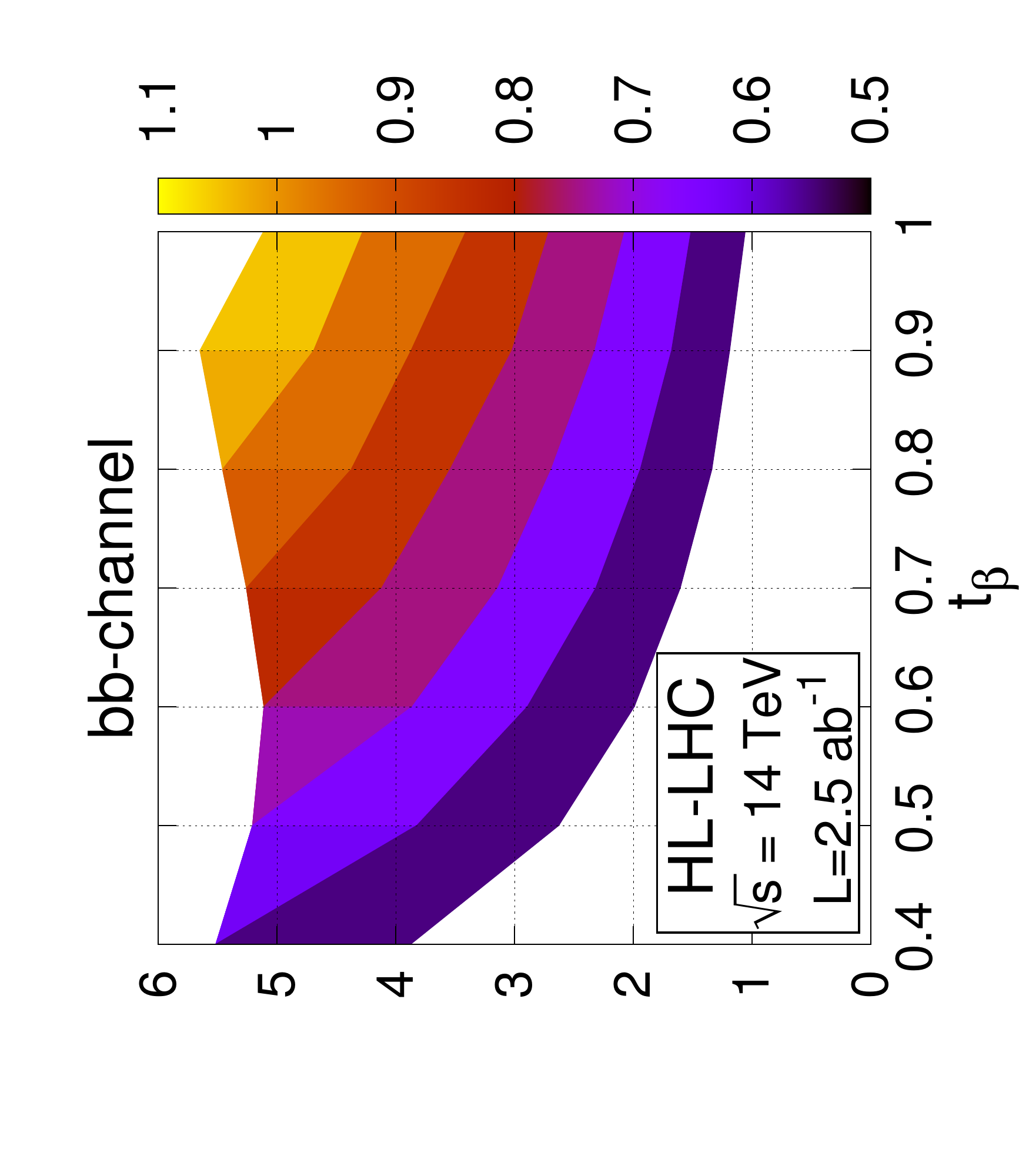}}
\subfigure[]{\includegraphics[scale = 0.24,angle=270]{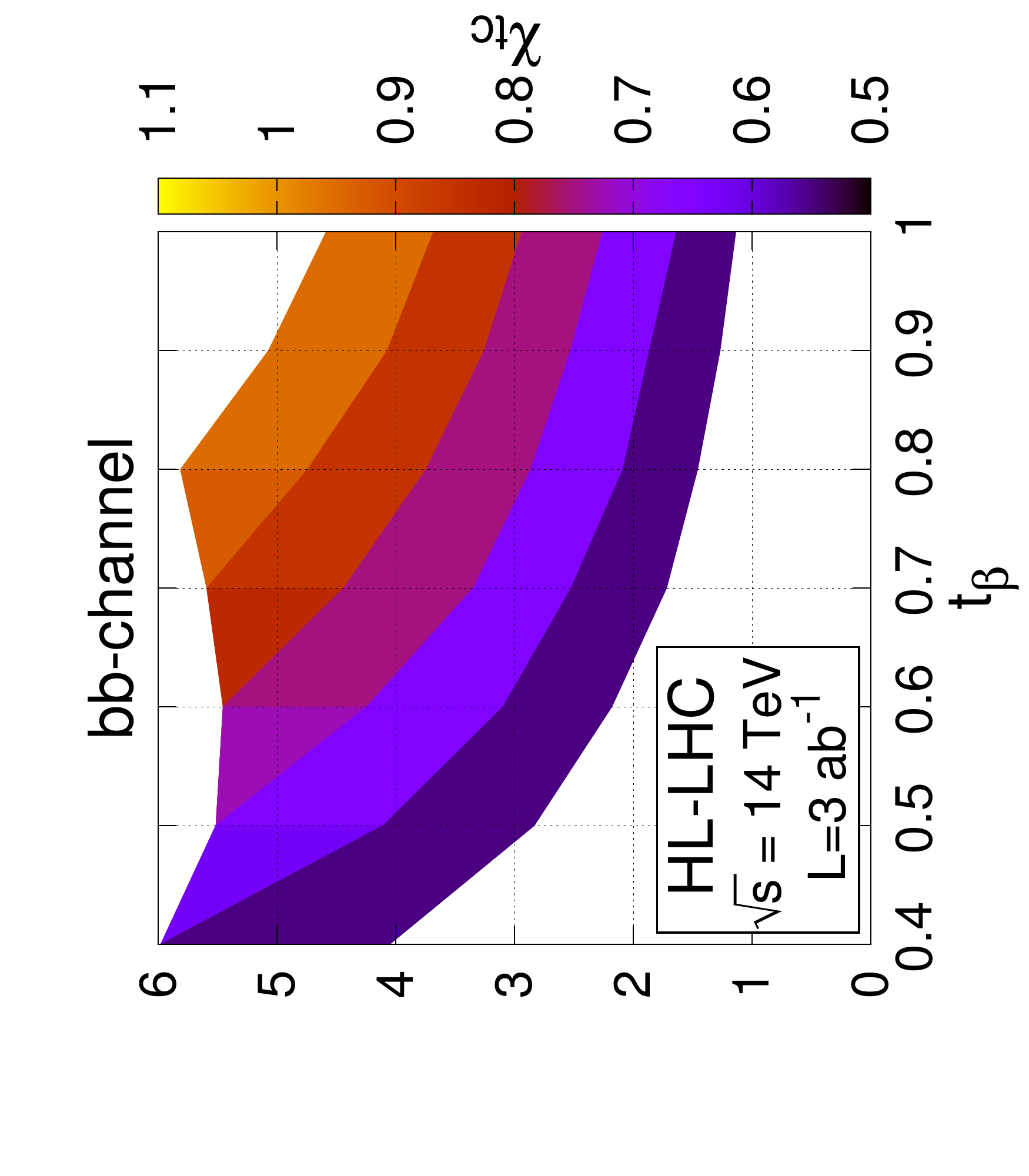}}
 \caption{Density plots for the signal significance as a function of the $t_{\beta}$ and $\chi_{tc}$ for three illustrative integrated luminosities: (a) $\mathcal{L}$=2 ab$^{-1}$, (b) $\mathcal{L}$=2.5 ab$^{-1}$, (c) $\mathcal{L}$=3 ab$^{-1}$. The case (c) represents the final integrated luminosity reached by the HL-LHC. \label{SigmaBBChannel_HL-LHC}}
	\end{figure}

\begin{figure}[!h]
\centering
\subfigure[]{\includegraphics[scale = 0.24,angle=270]{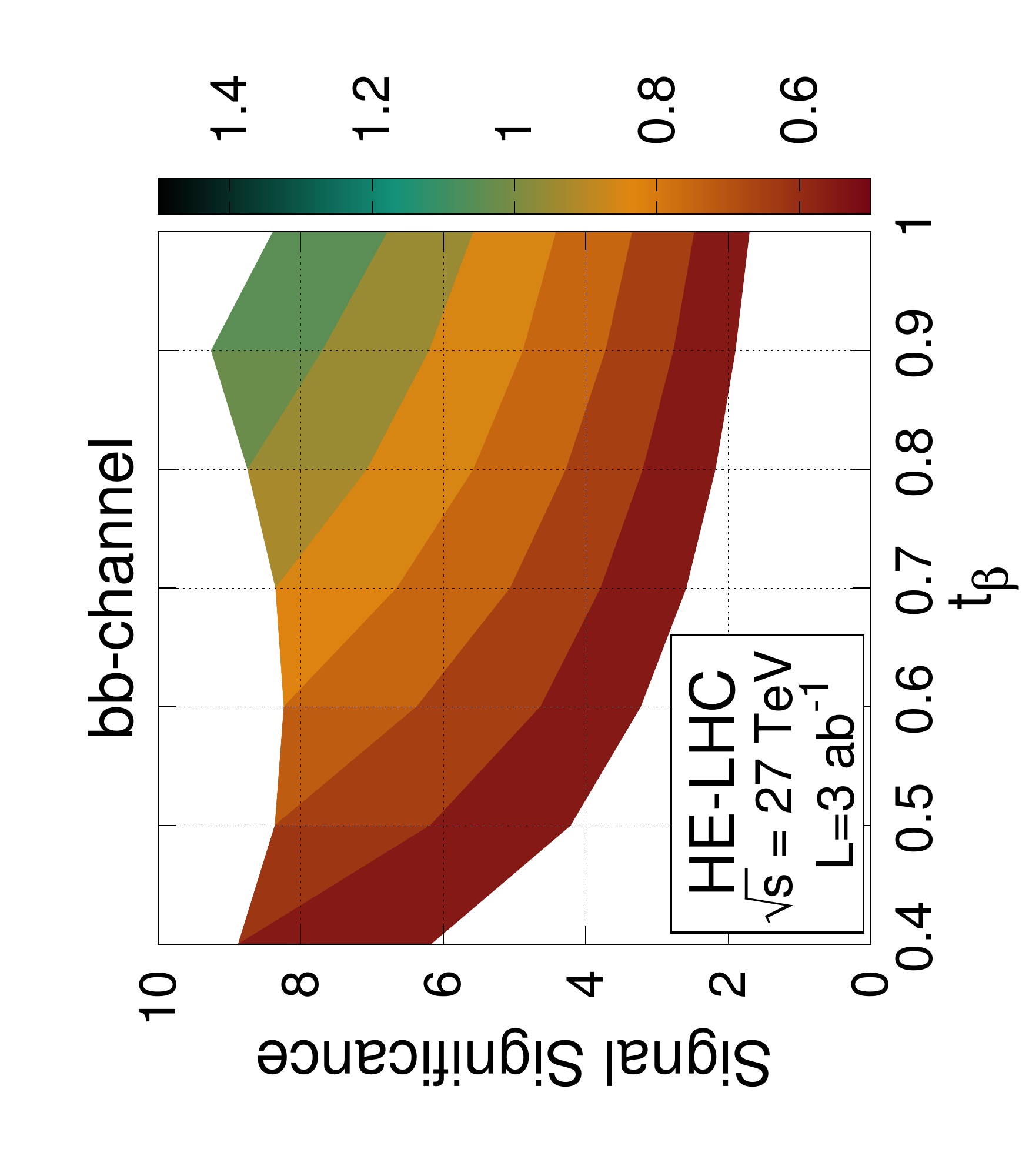}}
\subfigure[]{\includegraphics[scale = 0.24,angle=270]{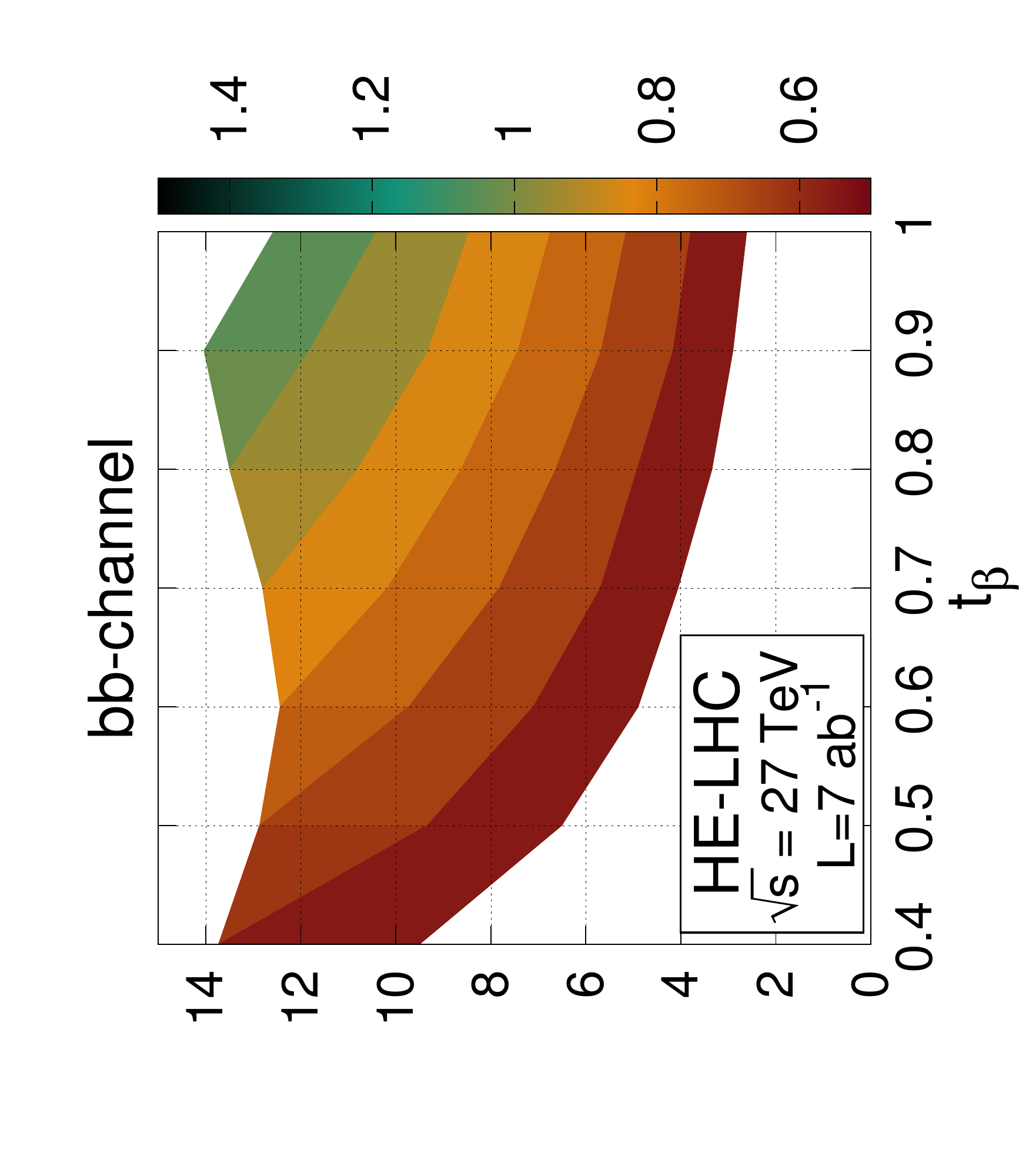}}
\subfigure[]{\includegraphics[scale = 0.24,angle=270]{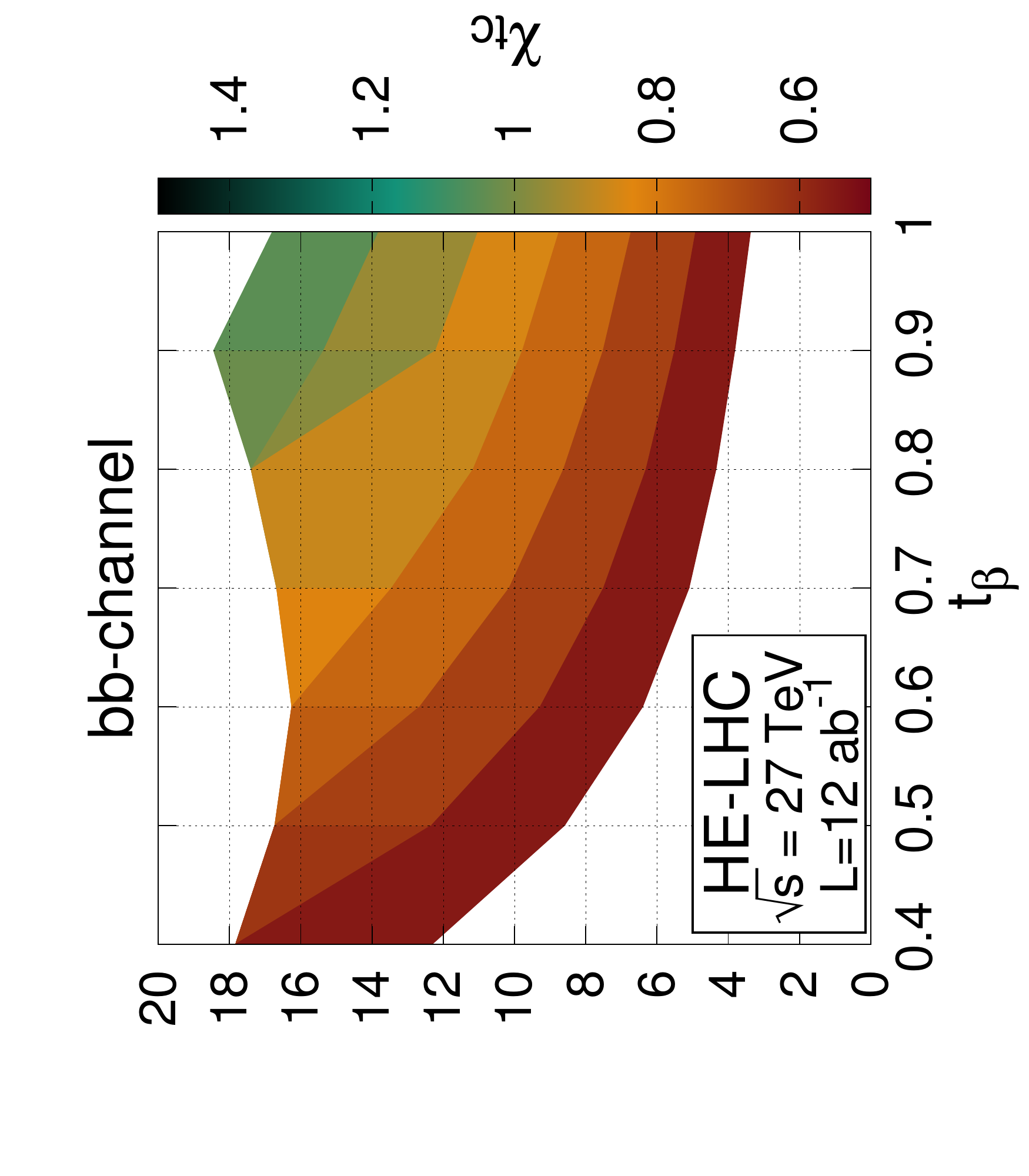}}
 \caption{The same as in figure \ref{SigmaBBChannel_HL-LHC} but for: (a) $\mathcal{L}$=3 ab$^{-1}$, (b) $\mathcal{L}$=7 ab$^{-1}$, (c) $\mathcal{L}$=12 ab$^{-1}$. The case (c) represents the final integrated luminosity reached by the HE-LHC. \label{SigmaBBChannel_HE-LHC}}
	\end{figure}

\begin{figure}[!h]
\centering
\subfigure[]{\includegraphics[scale = 0.24,angle=270]{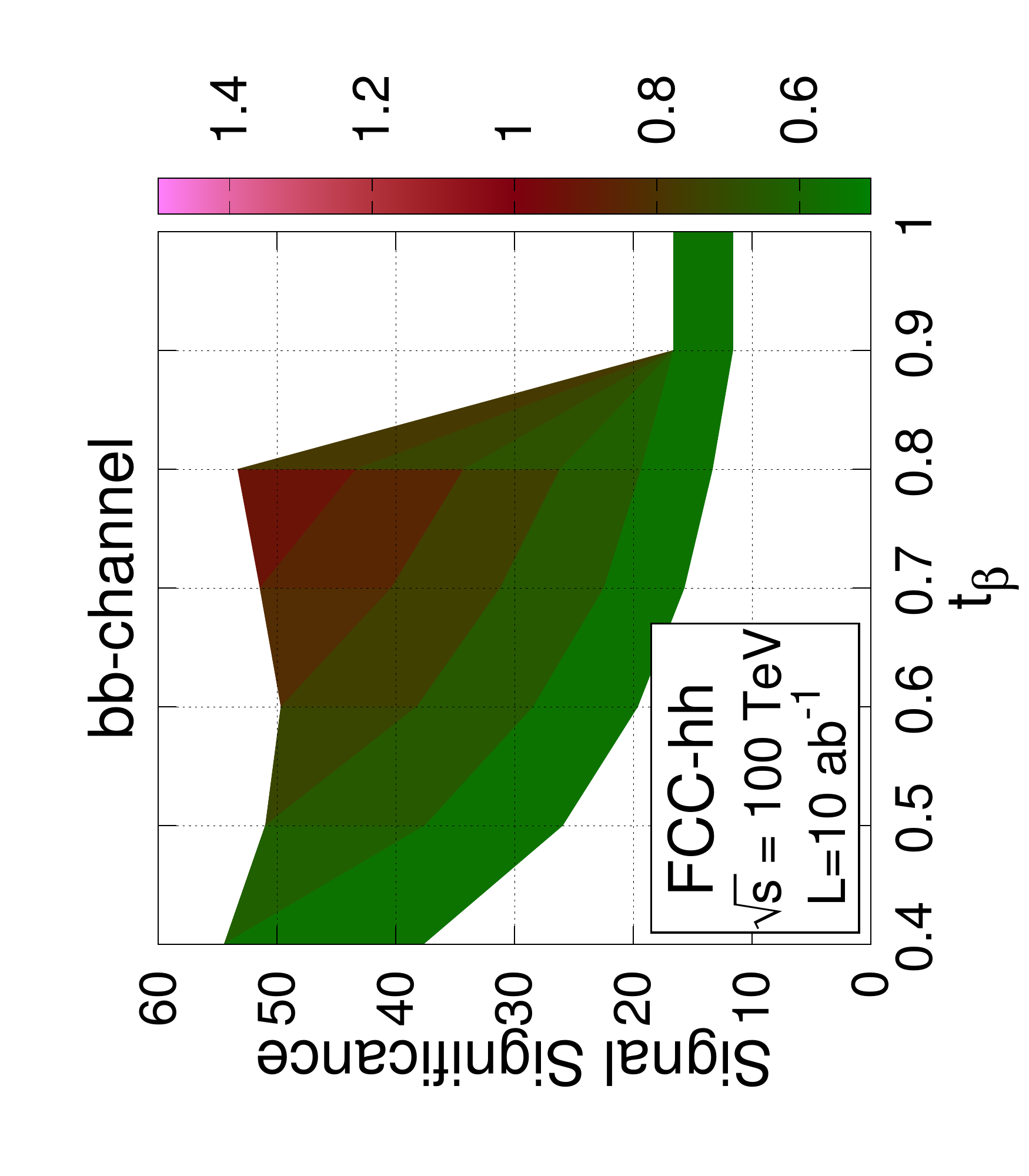}}
\subfigure[]{\includegraphics[scale = 0.24,angle=270]{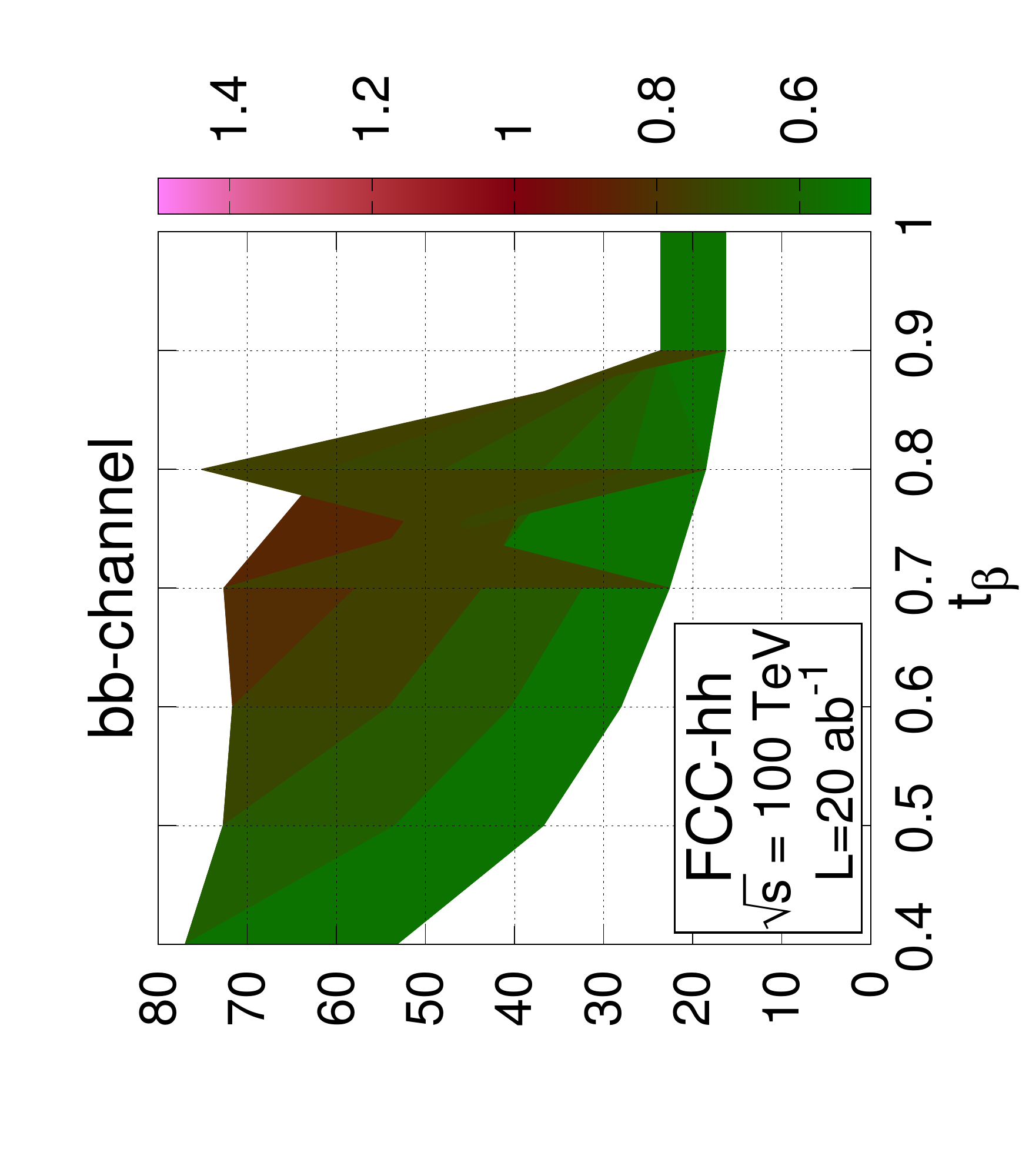}}
\subfigure[]{\includegraphics[scale = 0.24,angle=270]{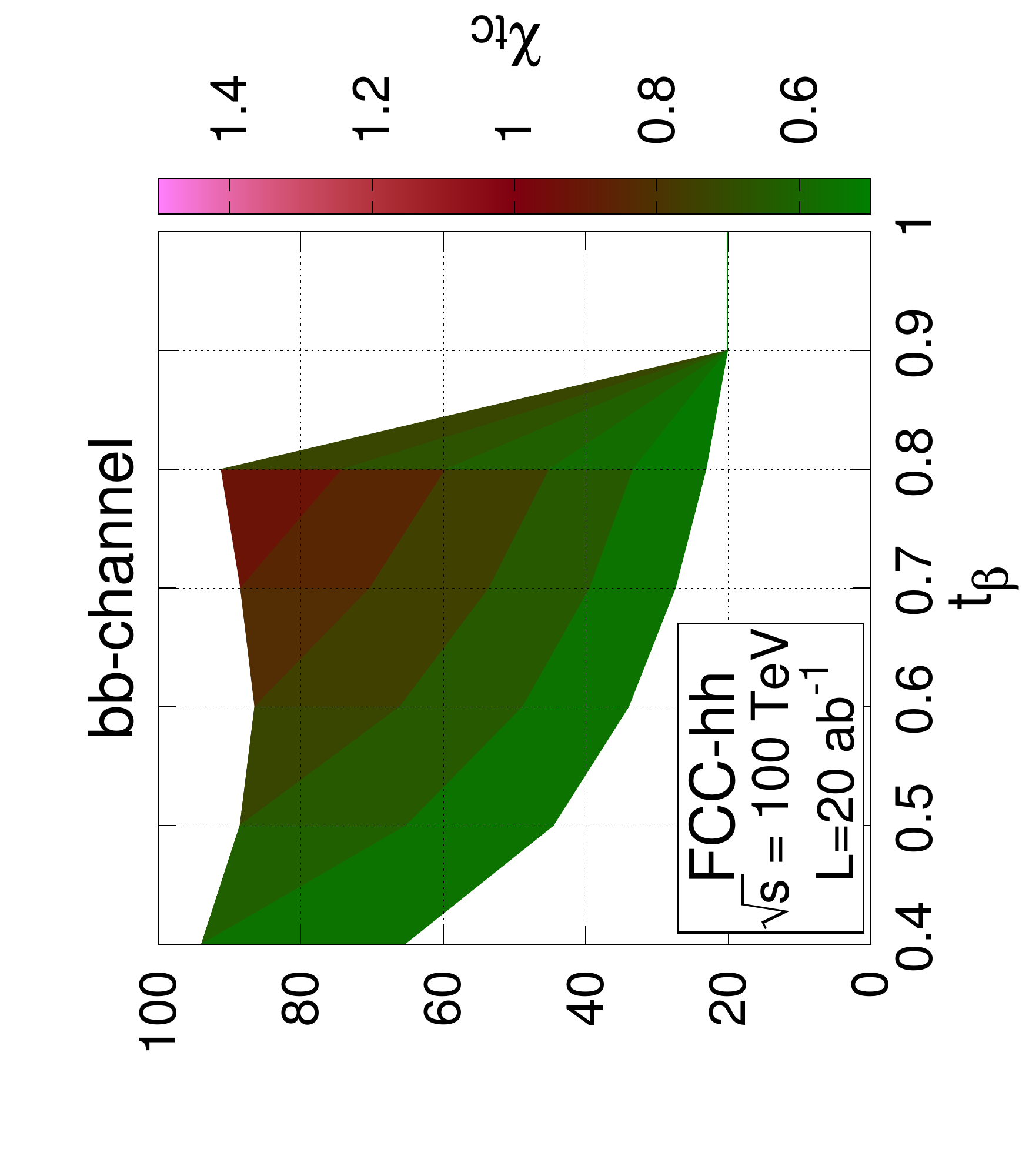}}
 \caption{The same as in figure \ref{SigmaBBChannel_HL-LHC} but for: (a) $\mathcal{L}$=10 ab$^{-1}$, (b) $\mathcal{L}$=20 ab$^{-1}$, (c) $\mathcal{L}$=30 ab$^{-1}$. The case (c) represents the final integrated luminosity reached by the FCC-hh. \label{SigmaBBChannel_FCC-hh}}
	\end{figure}

Once the kinematic cuts of the section \ref{cuts} are applied, luminosities larger than $\sim$500 fb$^{-1}$ are required to achieve a signal significance of $\sim 3\sigma$ at the LHC; although the HL-LHC is more promising. The figure \ref{SigmaBBChannel_HL-LHC} shows density plots for the signal significance as a function of $t_{\beta}$ and $\chi_{tc}$ for the HL-LHC by considering three values of the integrated luminosity, $\mathcal{L}=$2, 2.5, 3 ab$^{-1}$.
 The last value is the aim to search at the HL-LHC. Once the integrated luminosity exceeds a value of $\mathcal{L}\sim$2 ab$^{-1}$, a evidence for the $t\to ch$ decay could be claimed. With a luminosity of least 2.5 ab$^{-1}$, a potential discovery looks promising. Finally, when a luminosity of 3 ab$^{-1}$ is considered, it is the most encouraging scenario with up to $\sim 6\sigma$'s for ($t_{\beta}\sim$ 0.4, $\chi_{tc}\sim$ 0.5) and ($t_{\beta}\sim$ 0.8, $\chi_{tc}\sim$ 0.9). As far as to the HE-LHC and the FCC-hh are concerned, the results are even more promising than for the HL-LHC. The figure \ref{SigmaBBChannel_HE-LHC} and \ref{SigmaBBChannel_FCC-hh} presents density plots as the figure \ref{SigmaBBChannel_HL-LHC}, but for the HE-LHC and  the FCC-hh. 
Three representative scenarios, for both the HE-LHC and the FCC-hh, are explored also,  $\mathcal{L}=$3, 7, 12 ab$^{-1}$ and $\mathcal{L}=$10, 20, 30 ab$^{-1}$, respectively. Both colliders could be used to perform a cross-check since, for instance, at the HE-LHC with a minimum integrated luminosity of $0.5$ ab$^{-1}$ discovery of the $t\to ch$ decay could be announced. With  higher integrated luminosities, for instance, $\mathcal{L}=$12 ab$^{-1}$ and with ($t_{\beta}\sim$ 0.9, $\chi_{tc}\sim$ 1.1), a signal significance of $\sim 18\sigma$ is found. 
On the other hand, at the FCC-hh, signal significances of up to $\mathcal{O}$(90) are searched, with this values, the FCC-hh could work as a FCNC processes factory.

In the table \ref{table} we show a summary of the main results. 
\begin{table}[htb]
{\footnotesize{}\caption{Integrated luminosities for evidence or dicovery of the $t\to ch$
decay at hadron colliders.\label{table}}
}{\footnotesize \par}

\centering{}{\footnotesize{}}%
\begin{tabular}{cccc}
\hline 
\textbf{\footnotesize{}Collider} & \textbf{\footnotesize{}Energy} & \textbf{\footnotesize{}I. Luminosity for evidence (3$\sigma$)} & \textbf{\footnotesize{}I. Luminosity for discovery (5$\sigma$)}\tabularnewline
\hline 
\hline 
{\footnotesize{}LHC} & {\footnotesize{}14 TeV} & {\footnotesize{}No evidence} & {\footnotesize{}No discovery}\tabularnewline
\hline 
{\footnotesize{}HL-LHC} & {\footnotesize{}14 TeV} & {\footnotesize{}}%
\begin{tabular}{c}
\hline 
{\footnotesize{}$bb-channel:\;\sim0.5\, ab^{-1}$}\tabularnewline
\hline 
\hline 
{\footnotesize{}$\gamma\gamma-channel:\;\sim3\, ab^{-1}$}\tabularnewline
\hline 
\end{tabular} & {\footnotesize{}}%
\begin{tabular}{c}
\hline 
{\footnotesize{}$bb-channel:\;\sim2.5\, ab^{-1}$}\tabularnewline
\hline 
\hline 
{\footnotesize{}$\gamma\gamma-channel:$ NO}\tabularnewline
\hline 
\end{tabular}\tabularnewline
\hline 
{\footnotesize{}HE-LHC} & {\footnotesize{}27 TeV} & {\footnotesize{}}%
\begin{tabular}{c}
\hline 
{\footnotesize{}$bb-channel:\;\sim0.1\, ab^{-1}$}\tabularnewline
\hline 
\hline 
{\footnotesize{}$\gamma\gamma-channel:\;\sim0.3\, ab^{-1}$}\tabularnewline
\hline 
\end{tabular} & {\footnotesize{}}%
\begin{tabular}{c}
\hline 
{\footnotesize{}$bb-channel:\;\sim0.5\, ab^{-1}$}\tabularnewline
\hline 
\hline 
{\footnotesize{}$\gamma\gamma-channel:\;\sim1.7\, ab^{-1}$}\tabularnewline
\hline 
\end{tabular}\tabularnewline
\hline 
{\footnotesize{}FCC-hh} & {\footnotesize{}100 TeV} & {\footnotesize{}A few $fb^{-1}$} & {\footnotesize{}A few $fb^{-1}$}\tabularnewline
\hline 
\end{tabular}{\footnotesize{} }
\end{table}
{\footnotesize \par}


\newpage
\section{Conclusions\label{SeccionV}}
We study the $t\to ch$ decay at future hadron colliders,
namely, HL-LHC, HE-LHC and FCC-hh with center-of-mass energies associated
to each hadron collider, i.e, $\sqrt{s}$ = 14 (HL- LHC), 27 (HE-LHC)
and 100 TeV (FCC-hh). Integrated luminosities from $0.3$ to $30$
ab$^{-1}$ were explored. In this work we consider the Type-III Two-Higgs
Doublet Model for which two decay channels of the SM-like Higgs boson were
proposed and analized: into two photons $(\gamma\gamma-channel)$
and into two bottom quarks $(bb-channel)$. 
After studying the constraints
on the free model parameters from the most up-to-date Higgs boson
coupling and applying several kinematic cuts to the signal and SM
background, we find that with the integrated luminosity achieved at
the LHC, $0.3$ ab$^{-1}$, is not possible claim discovery for the
$t\to ch$ decay. 
However, in the $bb-channel$, an integrated luminosity
of at least $\sim0.5$ ab$^{-1}$ is necessary to achieve a signal
significance of $3\sigma$. On the other hand, with the forthcoming
HL-LHC, once it achieves an integrated luminosity of $\sim2.5$ ab$^{-1}$($\sim3$
ab$^{-1}$), discovery (evidence) in the $bb-channel$ $(\gamma\gamma-channel)$
could be claimed. More favorable results emerge for the HE-LHC since
with an integrated luminosity of $\sim0.5$ ab$^{-1}$ ($\sim1.7$
ab$^{-1}$), discovery of the $t\to ch$ decay in the $bb-channel$
$(\gamma\gamma-channel)$ will be announced. With these results, several
cross-checks, in both channels, could be performed. Finally, the most
promising scenario arises at the FCC-hh, which, among other goals,
could work as a FCNC factory rediscovering the $t\to ch$ decay with
a few fb$^{-1}$ of integrated luminosity in both channels.

\acknowledgments

We acknowledge support from CONACYT (M\'exico). The work of M. A. Arroyo-Ure\~na was supported by PAPIIT Project IN115319, DGAPA-UNAM. The authors thankfully acknowledge computer resources, technical advise and support provided by Laboratorio Nacional de Superc\'omputo del Sureste de M\'exico (LNS), a member of the CONACYT national laboratories, with project No. 201801027c.

\appendix
\section{Complementary formulas }
\subsection{Matrices of rotation \label{AppxAmatrices}}
The explicit form of the matrices that diagonalize the mass matrix, eq. \ref{matricesMASA}, are given by \cite{Fritzsch:1995nx}, \cite{Branco:1999nb}:
\begin{equation}\label{O}
\mathcal{O}_{f}=\left(\begin{array}{ccc}
\sqrt{\frac{m_{2}m_{3}(A-m_{1})}{A(m_{2}-m_{1})(m_{3}-m_{1})}} & \sqrt{\frac{m_{1}m_{3}(m_{2}-A)}{A(m_{2}-m_{1})(m_{3}-m_{2})}} & \sqrt{\frac{m_{1}m_{3}(A-m_{3})}{A(m_{3}-m_{1})(m_{3}-m_{2})}}\\
-\sqrt{\frac{m_{1}(m_{1}-A)}{(m_{2}-m_{1})(m_{3}-m_{1})}} & \sqrt{\frac{m_{2}(A-m_{2})}{(m_{2}-m_{1})(m_{3}-m_{2})}} & \sqrt{\frac{m_{3}(m_{2}-A)}{(m_{2}-m_{1})(m_{3}-m_{2})}}\\
\sqrt{\frac{m_{1}(A-m_{2})(A-m_{3})}{A(m_{2}-m_{1})(m_{3}-m_{1})}} & -\sqrt{\frac{m_{2}(A-m_{1})(m_{3}-A)}{A(m_{2}-m_{1})(m_{3}-m_{2})}} & \sqrt{\frac{m_{3}(A-m_{1})(A-m_{2})}{A(m_{3}-m_{1})(m_{3}-m_{2})}}
\end{array}\right),
\end{equation}
and
\begin{equation}\label{fases}
P_{f}=\left(\begin{array}{ccc}
1 & 0 & 0\\
0 & e^{i\alpha_{1}} & 0\\
0 & 0 & e^{i\alpha_{2}}
\end{array}\right),
\end{equation}
where $m_i$ $(i=1,\,2,\,3)$ are the physical fermion masses.
\subsection{Yukawa Lagrangian \label{AppxALagrangian}}
The Yukawa Lagrangian of the Type-III Two-Higgs Doublet Model in terms of the physical fields are given by:
\begin{eqnarray}
\mathcal{L}_{Y} & = & \frac{g}{2}\left(\frac{m_{d}}{m_{W}}\right)\bar{d}_{i}\left[\frac{\cos\alpha}{\cos\beta}\delta_{ij}+\frac{\sqrt{2}\sin(\alpha-\beta)}{g\cos\beta}\left(\frac{m_{W}}{m_{d}}\right)\left(\tilde{Y}_{2}^{d}\right)_{ij}\right]d_{j}H\nonumber \\
 & + & \frac{g}{2}\left(\frac{m_{d}}{m_{W}}\right)\bar{d}_{i}\left[-\frac{\sin\alpha}{\cos\beta}\delta_{ij}+\frac{\sqrt{2}\cos(\alpha-\beta)}{g\cos\beta}\left(\frac{m_{W}}{m_{d}}\right)\left(\tilde{Y}_{2}^{d}\right)_{ij}\right]d_{j}h\nonumber \\
 & + & i\frac{g}{2}\left(\frac{m_{d}}{m_{W}}\right)\bar{d}_{i}\left[-\tan\beta\delta_{ij}+\frac{\sqrt{2}}{g\cos\beta}\left(\frac{m_{W}}{m_{d}}\right)\left(\tilde{Y}_{2}^{d}\right)_{ij}\right]\gamma^{5}d_{j}A\nonumber \\
 & + & \frac{g}{2}\left(\frac{m_{u}}{m_{W}}\right)\bar{u}_{i}\left[\frac{\sin\alpha}{\sin\beta}\delta_{ij}+\frac{\sqrt{2}\sin(\alpha-\beta)}{g\sin\beta}\left(\frac{m_{W}}{m_{u}}\right)\left(\tilde{Y}_{2}^{u}\right)_{ij}\right]u_{j}H\\
 & + & \frac{g}{2}\left(\frac{m_{u}}{m_{W}}\right)\bar{u}_{i}\left[-\frac{\cos\alpha}{\sin\beta}\delta_{ij}+\frac{\sqrt{2}\cos(\alpha-\beta)}{g\sin\beta}\left(\frac{m_{W}}{m_{u}}\right)\left(\tilde{Y}_{2}^{u}\right)_{ij}\right]u_{j}h\nonumber \\
 &+&i\frac{g}{2}\left(\frac{m_{u}}{m_{W}}\right)\bar{u}_{i}\left[-\cot\beta\delta_{ij}+\frac{\sqrt{2}}{g\sin\beta}\left(\frac{m_{W}}{m_{u}}\right)\left(\tilde{Y}_{2}^{u}\right)_{ij}\right]\gamma^{5}u_{j}A,\nonumber
\end{eqnarray}
where $i$ and $j$ stand for the fermion flavors, in general $i\neq j$.
As far as the lepton interactions, it is similar to type-down quarks part with the exchange $d\to\ell$ and $m_d\to m_{\ell}$.


\end{document}